\documentclass[apj,revtex4]{emulateapj}
\usepackage{hyperref}

\hypersetup{
  colorlinks=true,
  urlcolor=blue,
  linkcolor=blue,
  citecolor=blue
}

\usepackage{graphicx}	% Including figure files
\usepackage{amsmath}	% Advanced maths commands
\usepackage{amssymb}	% Extra maths symbols

\def \AV{{$A_{\rm V}$}}
\def \RV{{$R_{\rm V}$}}

\def \msun{$\mathrm{M}_\odot$}
\def \rsun{$\mathrm{R}_\odot$}

\def \kms{km~s$^{-1}$}
\def \pap1{Paper I}
\def \pap2{Paper II}
\def \deg{$^\circ$}
\def \SST{$Spitzer~Space~Telescope$}
\def \na{New Astronomy}
\shorttitle{Eclipsing Binary Distance to Cyg OB2}
\shortauthors{Kiminki et al.}

\begin{document}

\title{Predicting {\it Gaia}'s Parallax Distance to the Cygnus OB2 Association with Eclipsing Binaries}

\author{Daniel C. Kiminki\altaffilmark{1}, Henry A. Kobulnicky\altaffilmark{2},
Carlos A. Vargas \'{A}lvarez\altaffilmark{2,3}, 
Michael J. Alexander\altaffilmark{2,4}, Michael J. Lundquist\altaffilmark{2}}
\altaffiltext{1}{Dept. of Astronomy, University of Arizona, 933 N. Cherry Avenue, Tucson, AZ 85721}
\email{kiminki@email.arizona.edu}
\altaffiltext{2}{Dept. of Physics \& Astronomy, University of Wyoming, 1000 E. University Avenue, Laramie, WY 82070}
\altaffiltext{3}{Physics Department, Hollins University, P.O. Box 9661, Roanoke, VA 24020}
\altaffiltext{4}{Dept. of Physics \& Astronomy, Lehigh University, 27 Memorial Drive West, Bethlehem, Pennsylvania 18015, USA }

\begin{abstract}
  The Cygnus OB2 Association is one of the nearest and largest
  collections of massive stars in the Galaxy.  Situated at the heart
  of the ``Cygnus X'' complex of star-forming regions and molecular
  clouds, its distance has proven elusive owing to the ambiguous
  nature of kinematic distances along this $\ell\simeq80\degr$
  sightline and the heavy, patchy extinction.  In an effort to refine
  the three-dimensional geometry of key Cygnus~X constituents, we have
  measured distances to four eclipsing double-lined OB-type
  spectroscopic binaries that are probable members of Cyg~OB2. We find
  distances of $1.33\pm0.17$, $1.32\pm0.07$, $1.44\pm0.18$, and
  $1.32\pm0.13$ kpc toward MT91~372, MT91~696, CPR2002~A36, and
  Schulte~3 respectively.  We adopt a weighted average distance of
  1.33$\pm$0.06~kpc.  This agrees well with spectrophotometric
  estimates for the Association as a whole and with parallax
  measurements of protostellar masers in the surrounding interstellar
  clouds, thereby linking the ongoing star formation in these clouds
  with Cyg~OB2.  We also identify Schulte 3C (O9.5V), a 4\arcsec\
  visual companion to the 4.75 day binary Schulte~3(A+B), as a
  previously unrecognized Association member.
\end{abstract}


%%%%%%%%%%%%%%%%%%%%%%%%%%%%%%%%%%%%%%%%%%%%%%%%%%%%%%%%%%%%%%%%%%%%%%%
\section{INTRODUCTION}

The Cygnus~OB2 Association (also known as VI Cygni) harbors the
largest number of O stars among nine OB associations within the
``Cygnus~X'' \citep{Piddington52} complex, and it is one of the
richest OB associations in the Galaxy.  Cygnus~X is a conspicuously
bright region in radio, X-ray, and infrared maps of the Galactic
Plane, covering nearly ten degrees near longitude $\ell$=80\degr\
\citep[review by][]{Odenwald93}. Discovered as an overdensity of
bright, reddened, early-type stars \citep{MunchMorgan53,
  JohnsonMorgan54, Schulte56, Schulte58}, Cyg~OB2 boasts $\sim$60 O
stars \citep{Wright15}, perhaps as many as $\sim$2600 B
stars \citep{Kn00}, and some of the most massive stars known
\citep{MT91,Comeron02}. These are all veiled by heavy and variable
interstellar extinction \citep[\AV = 4~--~12 mag;][]{TD91, MT91,
  Hanson03}. In recent years, large surveys have produced an
unprecedented panchromatic view of the interstellar environment that
produced Cyg~OB2.  These include a five-band mid-infrared survey using
the \SST\ \citep{Beerer10}, maps of the molecular clouds in several CO
transitions \citep{Schneider06,Leung92}, and maps of the dust
continuum at 1.2 mm \citep{Motte07}. O stars in Cyg~OB2 were the first
known stellar sources of X-ray emission \citep{Harnden79, Waldron98},
and several of the luminous members of the Association are
prototypical examples for whole classes of massive star phenomena,
such as the extremely luminous early-B supergiant Schulte 12
\citep[MT91~314 in the numeration of][]{MT91}, the non-thermal radio
source Schulte 9 \citep[MT91~431;][]{Abbott84, Naze12}, and the O3If*
star Schulte 7 \citep[MT457;][]{Walborn73}. As one of the nearest
regions of massive star formation, Cyg~OB2 also permits detailed
investigations of OB stars and their interplay with the interstellar
medium, such as the stellar bowshocks associated with high-velocity
members \citep{ComeronPasquali07, Gvaramadze07,Kobulnicky10}.

Despite its role as a nearby laboratory of massive star astrophysics,
the distance to Cyg~OB2 has proven elusive.  In part, this is the
result of variable extinction, which complicates spectrophotometric
distances. Additionally, its location is at $\ell=80^\circ$, where the
relation between radial velocity and distance is not only
double-valued but poorly-defined owing to velocity crowding along the
tangent point of the local spiral arm \citep[e.g.,
see][]{Dame85,Dame01}.  Spectrophotometric distance measurements span
a range that includes 1.5~kpc \citep{JohnsonMorgan54}, 2.1~kpc
\citep{RLP}, 1.7~kpc \citep{MT91}, 1.7~kpc \citep{TD91}, and 1.45~kpc
\citep{Hanson03}. Additionally, studies of the eclipsing, double-lined
spectroscopic binaries Cyg~OB2-B17 \citep{Stroud10} and V382~Cyg
\citep{Yasarsoy12} give distances of 1.5~--~1.8~kpc and
$1.466\pm0.076$~kpc respectively. This range places Cyg~OB2 within the
local spiral arm or spur but is insufficiently precise for measuring
key physical parameters such as luminosities and mass loss
rates. \citet{Rygl12} used radio-wave very-long-baseline
interferometry (VLBI) to measure parallaxes of five pre-main-sequence
water and OH masers within Cygnus~X and found that four of the five
had distances between 1.3 and 1.5~kpc with typical uncertainties of
0.1~kpc.  \citet{Zhang12} also used VLBI techniques to measure a
parallax distance of 1.61$\pm0.12$ kpc to the red hypergiant NML
Cygni, and they concluded that it may lie on the far side of the
Cyg~OB2 association. \citet{Dzib12} reported radio VLBA observations
giving a parallax distance of 1.65$^{+0.96}_{-0.44}$~kpc to the
colliding-wind binary Schulte 5, and they infer a probable distance of
1.3~--~1.4 kpc based on estimates of the orbital parameters in this
quadruple system. While these parallax determinations are the most
precise distances yet published, Cygnus~X probably has a significant
spread along the line of sight. Because of this, individual OB
associations and star forming regions may lie anywhere within this
depth. Thus, independent distance measurements of individual features
are needed to help refine the line of sight structure of this complex,
in particular for Cyg~OB2.

In this contribution, we use four eclipsing double-lined spectroscopic
binaries (SB2s) investigated as part of the Cygnus~OB2 Radial Velocity
Survey to calculate an independent distance to the Association. Papers
I~--~VII in this series \citep{Kiminki07, Kiminki08, Kiminki09,
  Kiminki12a, Kiminki12b, Kobulnicky12, Kobulnicky14} describe prior
results of the Cygnus~OB2 Radial Velocity Survey which measured the
massive binary characteristics (i.e., binary fraction, distribution of
periods, mass ratios, eccentricities) for more than 100 massive stars
in a single cluster/association having a common formation environment.
Paper V \citep{Kiminki12b}, in particular, uses these basic data to
infer the intrinsic distributions of massive binaries, concluding that:
the fraction of massive stars having companions may be as high as
90\%, 45\% of these multiples are likely to interact at some point,
there exists an excess of short-period 4~--~7 day systems relative to
7~--~14 day systems, and that unresolved secondaries contribute
$\sim$16\% of the light in young stellar populations. Paper VII uses
the extensive repository of orbital parameter information for the 48
known massive binaries to show that no single power law provides a
statistically compelling prescription of the cumulative orbital period
distribution, and that a flattening of the distribution at $P>45$~d
indicates either a lower binary fraction or a shift toward low-mass
companions among long-period systems.

Our present dataset includes optical spectra and broadband photometry
of four eclipsing double-lined spectroscopic binaries: MT91~372 and
MT91~696 \citep[nomenclature of][]{MT91}, CPR2002~A36
\citep[nomenclature of][also known as RLP357 in
\citealt{RLP}]{Comeron02}, and Schulte~3 \citep[nomenclature of][also
known as RLP920 in \citealt{RLP}]{Schulte58}. Additionally, we make
use of the Northern Sky Variability Survey (NSVS) photometry
\citep{Wozniak2004} for CPR2002~A36 and Schulte~3. In Paper III
\citep{Kiminki09}, we presented a single-lined orbital solution and
NSVS photometry with a period of 2.22~days for MT91~372 (B2?+early
B).  \citet{Rios04} reported the SB2 (O9.5V+early B) nature of
MT91~696 and estimated a photometric period of 1.46 d. A refined
photometric period of $1.46919\pm0.00006$~d appears in
\citet{Souza14}. We presented the first spectroscopic orbital solution
in Paper IV with a similar period \citep{Kiminki12a}.  \citet{NSVSa}
identify the eclipsing nature of CPR2002~A36 using NSVS photometry.
Its SB2 nature was noticed by \citet{Hanson03}, and we presented the
orbital solution for this P=4.67~d system in Paper~III
\citep{Kiminki09}.  Schulte~3 was discovered to be a double-lined
(O6IV+O9III) spectroscopic binary with P=4.74 d in Paper II of this
series \citep{Kiminki08}.  In this work, we analyze the light curve of
this eclipsing system for the first time.

Section 2 of this work describes the new photometric and spectroscopic
data obtained for the purpose of creating joint velocity curves and
light curves for these four eclipsing double-lined systems.  Section 3
reports the discovery of additional visual companions at small angular
separations from Schulte 3 and MT91~696; knowledge of these
third-light contributions is essential for accurate luminosities and
distances.  Section 4 provides details of the combined
light-curve/velocity-curve modeling used to infer the distance to each
system.  Section 5 summarizes the new eclipsing binary distance to
Cyg~OB2 in relation to measurements of other targets in this
region. All radial velocities reported here are in the Heliocentric
reference frame.

%%%%%%%%%%%%%%%%%%%%%%%%%%%%%%%%%%%%%%%%%%%%%%%%%%%%%%%%%%%%%%%%%%%%%%%
\section{OBSERVATIONS AND DATA REDUCTION}

\subsection{Spectroscopic Data}

Optical spectroscopic observations of the four targets were obtained
as part of the Cygnus~OB2 Radial Velocity Survey on numerous dates
between 2001 and 2011 using a variety of telescopes and instruments as
described in Papers I~--~IV of this series.  Several additional
observations were obtained in 2012 October (on MT91~696 and Schulte~3)
and 2013 May~--~June (on MT91~372) using the Wyoming Infrared
Observatory (WIRO) Longslit spectrograph with an e2V 2048$^2$ CCD as
the detector.  A 2000~l~mm$^{-1}$ grating in first order yielded a
spectral resolution of 1.25~\AA\ near 5800~\AA\ with a
1\farcs2$\times$100\arcsec\ slit.  The spectral coverage was
5250~--~6750~\AA. Exposure times ranged between 1200~s and 2 hours in
multiples of 600~s depending on target brightness, current seeing
(1\farcs2~--~3\arcsec\ FWHM) and cloud conditions.  Reductions followed
standard longslit techniques, including flat fielding from dome quartz
lamp exposures. Copper-argon arc lamp exposures were taken after each
star exposure to wavelength calibrate the spectra to an rms of
0.03~\AA\ (1.5~\kms\ at 5800~\AA). Multiple exposures were combined
yielding final signal-to-noise ratios (SNR) typically in excess of
100:1 near 5800~\AA. Final spectra were Doppler corrected to the
Heliocentric velocity frame.  Each spectrum was then shifted by a
small additional amount in velocity so that the \ion{Na}{1}~D
$\lambda\lambda$5890,5996 lines were registered with the mean
\ion{Na}{1} line wavelength across the ensemble of observations. This
zero-point correction to each observation is needed to account for
effects of image wander in the dispersion direction when the FWHM of
the stellar point spread function was less than the slit width.
Because of these inevitable slit-placement effects on the resulting
wavelength solutions, radial velocity (RV) standards were not
observed.  Relative velocity shifts were generally less than 6~\kms,
comparable to the magnitude of the random velocity uncertainties.

We measured an initial (final for CPR2002~A36) radial velocity for
each spectrum obtained at WIRO using Gaussian fits to the
\ion{He}{1}~$\lambda$5876 line via the same method outlined in
\citet{Kobulnicky12}, adopting a rest wavelength of 5875.69~\AA\
measured in model stellar spectra with static atmospheres
\citep[TLUSTY;][]{Lanz,Hubeny}.  Our fitting code\footnote{We use the
  robust curve fitting algorithm MPFIT as implemented in IDL
  \citep{mpfit}.} fixes the Gaussian width and depth at the mean value
determined from all the spectra (after rejecting outliers), and it
solves for the best-fit line center.  In the case of an SB2, the code
fits a double-Gaussian profile where the widths and depths have been
fixed independently using observations obtained near quadrature
orbital phases.

In this work, we used the initial radial velocities to disentangle the
component spectra of MT91~372, MT91~696, and Schulte~3 using the
method of \citet{GL2006}.\footnote{Disentangling is not performed for
  CPR2002~A36 on account of having an insufficient number of
  high-quality spectra.}  One of the strengths of this method is that
the radial velocities can be refined via cross correlation after each
iteration (i.e, cross-correlating the residual spectra with the
resultant component spectrum as the template). The cross-correlated
velocities generally have smaller uncertainties and utilize more
lines, leading to a more precise measurement of the stellar velocity
and the computed binary systemic velocity. Systemic velocities that
are based solely on one hydrogen or helium line, such as
\ion{He}{1}~$\lambda$5876 with CPR2002~A36, may be blueshifted
relative to the true systemic velocity owing to velocity contributions
from strong winds. These contributions are strongest in very early-O
stars and evolved stars. However, radial velocities for Schulte~3 are
measured from many lines in the blue and red portions of the
spectrum. The radial velocities of MT91~372 and MT91~696 are based on
several hydrogen and helium lines between 5400~\AA\ and 6700~\AA\, and
the stars are neither early-O or evolved.

Because of the more sophisticated approach to measuring the radial
velocities in this work, the newly determined velocities will vary
slightly from our previous works. This will be most evident for
Schulte~3 and CPR2002~A36, which now have smaller O-C velocity
residuals, indicating that a superior orbital solution has been
obtained with the use of new data and
methods. Table~\ref{specdata.tab} in the Appendix lists the
Heliocentric Julian dates (column~1), orbital phase (column~2), radial
velocities and uncertainties (columns~3 and 5), and the O-C residuals
(columns~4 and 6) of each component for each of the four eclipsing SB2
systems.

\subsection{Photometric Data}

Photometry of MT91~696 was obtained on nine nights between 2012
October 17 and November 10 at the University of Wyoming 0.6 meter Red
Buttes Observatory (RBO) using a 1024$\times$1024 Apogee Alta U47 CCD
through a $V$ filter.  Data on Schulte~3 was obtained on 14 nights at
RBO between 2012 October 17 and 2012 December 21, and data on
CPR2002~A36 was obtained on four nights between 2012 December 10 and
2012 December 22. The 0\farcs54 pixel$^{-1}$ scale at RBO yielded a
9.0\arcmin\ field of view centered on each object.  Photometry of
MT91~372 was also obtained at RBO on seven nights between 2013 June 14
and 2013 July 8 using an Alta U16 4096$\times$4096 CCD binned
2$\times$2, yielding 0\farcs73 pixels over a 24\arcmin\ field of view.
The observations span 10 orbital periods of MT91~372, 16 orbital
periods of MT91~696, two orbital phases of CPR2002~A36, and 13 orbital
periods of Schulte~3. Seeing varied between 3\arcsec\ and 5\arcsec\
FWHM.  Sky conditions were predominantly non-photometric. Sequences of
180~s, 120~s, and 30~s exposures were obtained for all systems, for
several hours each night.  Photometric conditions on 2012 October 18
(local) allowed us to obtain calibration images of several Landolt
standard fields and fields in Cyg~OB2 having {\it UBV} photometry in
\citealt[][(MT91)]{MT91}. We elected to perform differential
photometry using OB stars measured by \citet{MT91} in the $V$-band
because of the lack of suitably red ($B-V>1.2$) Landolt standards
required to match the colors of the heavily reddened OB star targets.
Additionally, we observed each target in the {\it UBV} filters on the
(mostly) photometric night of 2012 December 09, interleaving target
fields with nearby Cyg~OB2 fields for purposes of tying our photometry
to the MT91 {\it UBV} measurements.

RBO images were reduced using standard procedures which included
removing the zero level with bias exposures, removing dark current
using a median of at least seven dark exposures scaled to the
appropriate target exposure time, and dividing by a normalized flat
field constructed from the median of at least seven twilight sky
exposures having a minimum of 10,000 electrons per image. We performed
aperture photometry using IRAF/PHOT with an aperture of 14 pixels
(8\arcsec) and a sky annulus extending from 14 to 19 pixels, in order
to extract $>99$\%\ of the stellar flux for the target and 6~--~9
reference stars in each field. The magnitudes of the target stars were
compared to the average of the comparison stars in each field to
correct for (sometimes large) variations in atmospheric transmission.
Differential photometry of MT91~372 used nine nearby field stars for
reference, including four early type Cyg~OB2 stars having similar
color.  The zero point of the photometric calibration relies upon
stars from MT91 in the same field. The magnitudes of each comparison
star were examined relative to the mean for the remainder of the
stars, and in this way, we judge that the comparison stars were
non-variable to the level of 0.015 mag over the course of our program.
Photometric errors of the target star lie in the range 0.001~--~0.005
mag when considering only photon statistics as the dominant source of
uncertainty. The exposure-to-exposure rms suggests that this is an
appropriate estimate of the relative photometric uncertainty on any
given night once brief time periods of heavy cloud cover were excised
from the data.  However, the rms difference of the comparison star
mean magnitudes over all observing nights suggests that the
night-to-night zero point is uncertain at the 0.010 mag level, which
is small compared to the depth of eclipses over the orbital cycle.

For Schulte~3 and CPR2002~A36, we included the NSVS photometry
\citep{Wozniak2004} in addition to the RBO photometry. The NSVS
photometry were taken with four Apogee AP-10 imaging cameras with an
unfiltered optical response of $\sim$450~--~1000~nm and an effective
wavelength close to Johnson R-band. The data were collected over the
course of one year, from 1999 April 1 through 2000 March 30 and
exhibit an average photometric uncertainty of 0.01~mag. The NSVS data
used here are converted from Modified-Julian Date (MJD) to Julian Date
(JD).

Tables~~\ref{MT372phot.tab}~--~\ref{S3phot.tab} in the Appendix list
the Heliocentric Julian date and the $V$-band magnitudes of the
MT91~372, MT91~696, CPR2002~A36, and Schulte~3 systems. Column 1 gives
the Heliocentric Julian dates minus 2,400,000, while columns 2 and 3
list the $V$-band magnitudes and 1$\sigma$ uncertainties calculated
from photon statistics only.  Additional uncertainties at the level of
0.010 mag are present as a result of uncorrected atmospheric
transparency variations over the field of view and night-to-night zero
point uncertainties.  Only the first ten table lines are given to
provide context and format. The full table appears in the
\emph{electronic edition} as a machine-readable table.  The Schulte~3
photometry includes the contribution from the blended tertiary
Schulte~3C discussed in the next section. Additionally, the data have
been averaged into 2-minute intervals from the original 30-second
exposure times. MT91~696 photometry also includes a contribution from
the blended tertiary MT91~696C, discussed in the next section.

\section{Visual Companions to Schulte~3 and MT91 696}

\subsection{Schulte~3C}  

\begin{figure}[htpb!]
\centering
\includegraphics[width=0.95\columnwidth]{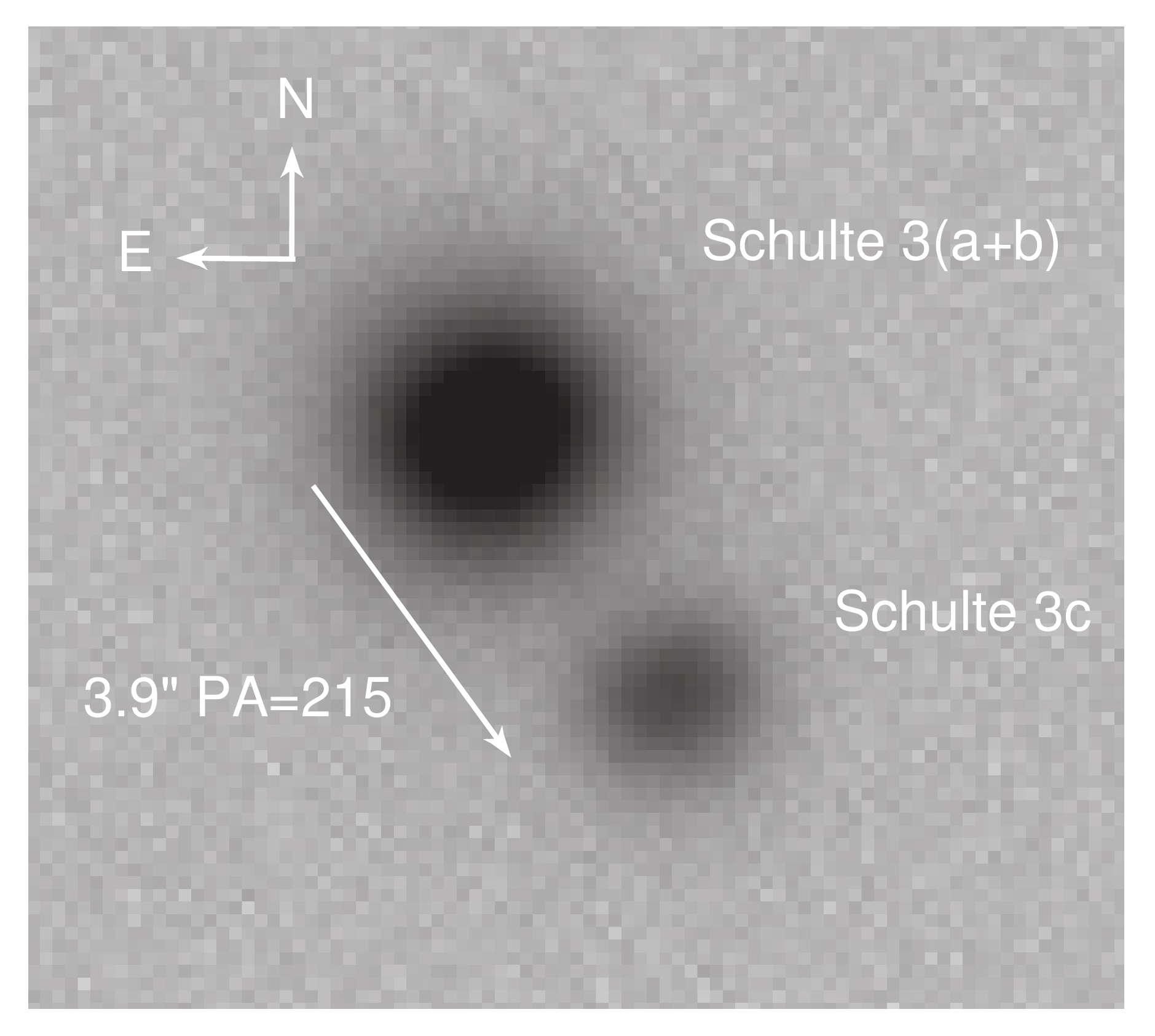}
\caption{Greyscale image of Schulte~3 (the eclipsing double-lined
  components a+b) and its $\sim$O9.5V companion,
  Schulte~3C. \label{S3img}}
\end{figure}

Photometry of the double-lined eclipsing binary Schulte~3 was
complicated by the presence of a nearby companion at 3\farcs9
separation at position angle 215\degr\ from the brightest visual
component.  Schulte~3 (comprised of 4.74~d eclipsing components ``A''
and ``B'') and the companion, Schulte~3C, are blended in all of our
RBO images, so our photometry using large apertures encompasses all
stars. The companion is readily visible in UKIDSS Galactic Plane
Survey images \citep{Lucas08} at R.A.=20:31:37.31
Declination=+41:13:17. (J2000), but both Schulte~3(A+B) and Schulte~3C
are saturated.  The companion is not present in the 2MASS Point Source
Catalog because of blending. It is not visible in the Digitized Sky
Survey for the same reason.  We therefore conducted our own $V$-band
imaging using the CCD guide camera on the WIRO-Longslit spectrograph
(0\farcs147 pixel$^{-1}$) on the night of 2012 November 06. Several
3 s exposures were obtained in 1\farcs4 seeing within a few minutes
of JD=2456238.55 (2012 Nov 7, 01:19 UT).  Figure~\ref{S3img} shows a
greyscale representation of our $V$-band images, illustrating the
relative locations of Schulte~3(A+B) and Schulte~3C.  Photometry using
these images indicates a magnitude difference $\Delta m_V$=2.14 mag,
meaning that Schulte~3(A+B) contribute 88\% of the flux at $V$ band,
compared to 12\% for Schulte~3C.  Given a mean magnitude of
V=10.30$\pm$0.02 on this date, derived from the phased light curve of
Schulte~3 (\S4.5), the companion has V$\simeq$12.42$\pm$0.03. It
also means the thirdlight-adjusted $V$-band maximum (between primary
and secondary eclipse) for the Schulte~3(A+B) system, as derived from
the phased light curve, is 10.31 mag.

\begin{figure}[htpb!]
\centering
\includegraphics[width=\columnwidth]{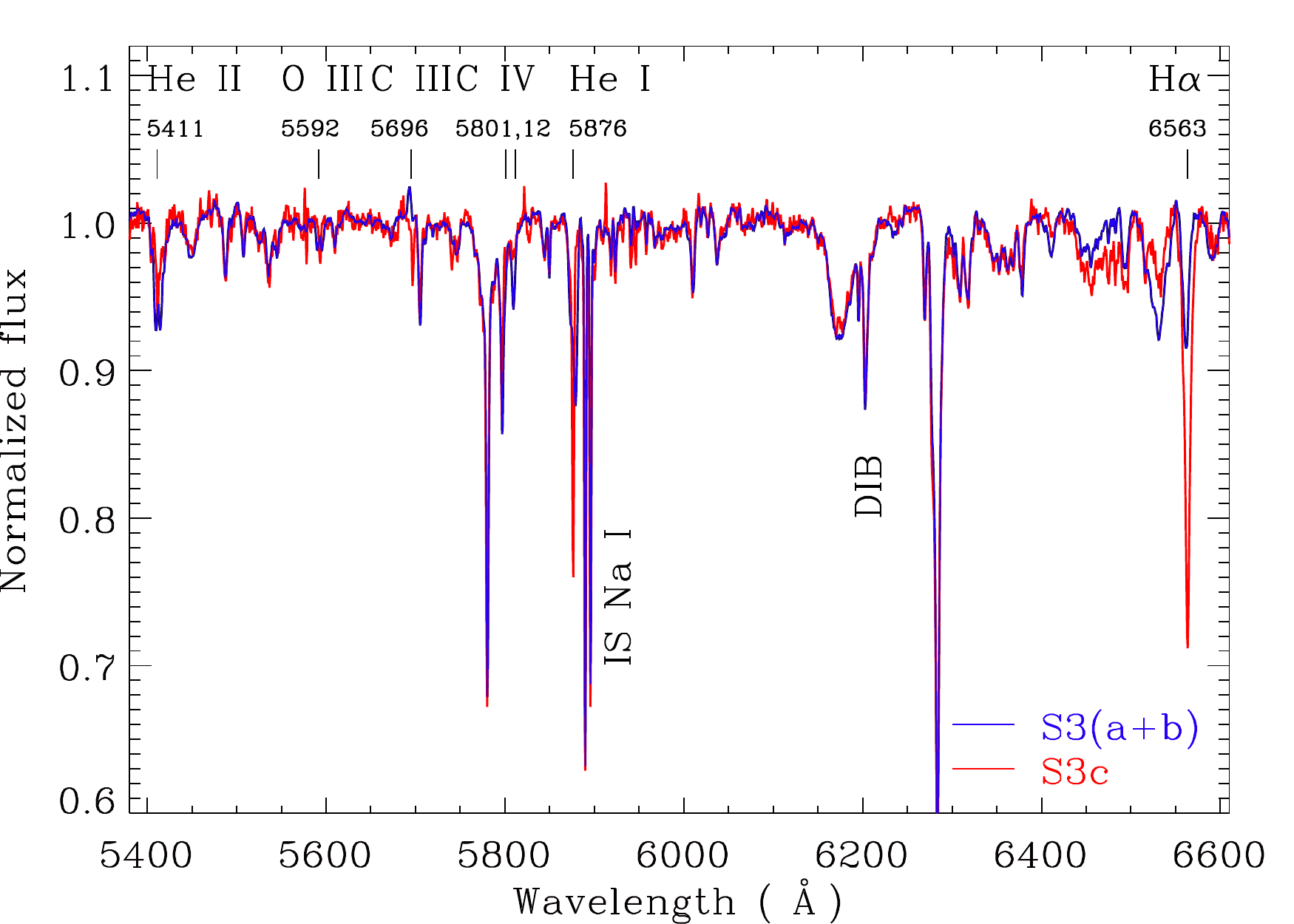}
\caption{Spectra of the O6V+O9III double-lined system Schulte~3(A+B)
  (blue) and its 4\arcsec\ O9.5V companion, Schulte~3C
  (red).  Labels mark key stellar features.  Other features are
  interstellar. The excellent agreement between the interstellar
  features of both stars are consistent with a common distance.  The
  two stars differ most notably in their \ion{He}{1} $\lambda$5876 and
  $H\alpha$ line depths, reflecting the higher effective temperature
  of Schulte~3(A+B). \label{S3spec}}
\end{figure}

We obtained spectra of Schulte~3(A+B) and Schulte~3C using the WIRO
Longslit spectrograph on the night of 2012 October 21 in 2\arcsec\
seeing. Figure~\ref{S3spec} plots the spectra of both components and
labels some key spectral features of hot stars.  The spectrum of
Schulte~3C exhibits both \ion{He}{2} $\lambda$5411 and \ion{He}{1}
$\lambda$5876 in the ratio EW$_{5411}$/EW$_{5876}\sim$ 0.4, consistent
with a temperature class near O9.5 \citep{Kobulnicky12b}.  H$\alpha$
appears in absorption.  The combination of the spectral
characteristics and the $V$-band magnitude are both consistent with an
O9.5V at the distance of Cyg~OB2. With the current data it is not
possible to say whether the companion is physically associated with
Schulte~3, although the strong similarity of the many diffuse
interstellar features in the two spectra make it likely that they both
lie at a similar distance. If associated, Schulte~3 is a triple
stellar system with Schulte 3C at a projected separation of about 5200
AU.

In the analysis that follows, we assume, but do not have the data to
demonstrate, that Schulte~3C is non-variable, both photometrically and
spectroscopically.  We are confident that it cannot be the source of
the photometric variability, detailed in the next section because the
spectroscopic period of Schulte~3(A+B), as observed in isolation,
matches the photometric period observed for the whole system.
However, the present data do not rule out the possibility that
Schulte~3C may be a multiple subsystem or may be photometrically
variable.

\subsection{MT91 696C and 696D}  

While no similarly bright companions are seen in UKIDSS images for
MT91~696, measurements with the {\it Hubble Space Telescope} Fine
Guidance Sensors (FGS) \citep{Caballero2014} reveal two astrometric
companions at 0\farcs023 (MT91~696C) and 0\farcs84\ (MT91~696D)
separations.  A Cyg~OB2 distance of 1.33~kpc implies minimum projected
physical separations of 31~AU and 1117~AU respectively, indicating
that this is likely a triple (possibly quadruple) system composed of a
spatially unresolved 1.4 day binary (components A \&\ B) and two
additional components (C \&\ D). With $\Delta$m=0.94$\pm$0.40 mag in
the $HST$ F583W bandpass, MT91~696C is the dominant source of excess
light (MT91~696D is several magnitudes fainter), so we consider only
this component in subsequent analysis of the light curve. The
magnitude of MT91~696C would make it a probable mid-B star. As with
Schulte~3, we assume, but do not have the data to demonstrate, that
MT91~696C is non-variable, both photometrically and spectroscopically.
It is also not the source of photometric variability for the same
reasons as Schulte~3C. Lastly, while we consider photometric
contamination from MT91~696C, we assume negligible radial velocity
contamination given a computed semi-amplitude of $\sim$5.5~\kms and
total flux contribution of $\sim$14\%\ (see Section~4.4).

\section{Analysis of Eclipsing Binaries}
\subsection{Modeling the Joint Light Curves and Velocity Curves}

\begin{figure}[htpb!]
\centering
\includegraphics[width=\columnwidth,trim=10mm 0 0 0]{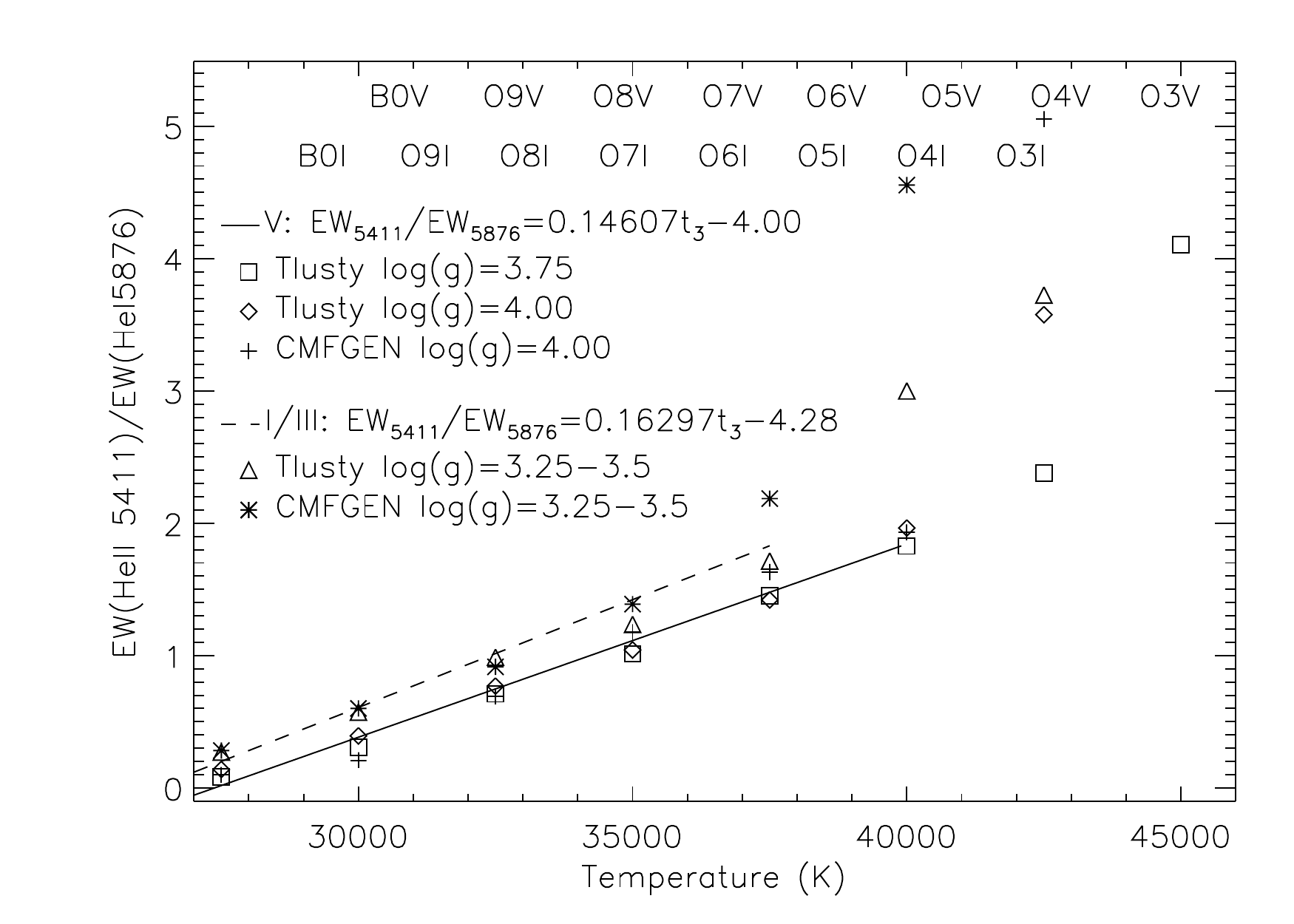}
\caption{Relation between equivalent width ratio \ion{He}{2}
  $\lambda$5411/\ion{He}{1} $\lambda$5876 versus temperature based on
  Tlusty \citet{Lanz} and CMFGEN \citep{Hillier98} NLTE model
  atmospheres for values of $\log$($g$) appropriate to main-sequence
  (3.75~--~4.0) and giant/supergiant (3.25~--~3.5) late-O and early B
  stars. The expressions within the panel quantify this quasi-linear
  relationship over the range 27,500~K~--~37,500~K for luminosity
  class V stars (solid line) and 27,500~K~--~35,000~K for luminosity
  class I/III stars (dashed line), where $t_3$ is temperature in
  10$^3$ K. \label{EWHe}}
\end{figure}

In general, effective temperatures for the primary star, $T_1$, were
held as fixed parameters based on the ratio of equivalent widths
\ion{He}{2}~$\lambda$5411/\ion{He}{1}~$\lambda$5876 in our spectra.
Figure~\ref{EWHe} plots this EW ratio versus temperature as measured
in Tlusty NLTE model atmosphere spectra \citep{Lanz} and CMFGEN model
spectra \citep{Hillier98}\footnote{As measured from the grid of O star
  stellar models dated 2009 June 24 from the webpage of J. Hillier,
  http://kookaburra.phyast.pitt.edu/hillier/web/CMFGEN.htm.} for four
values of $\log$($g$), 4.0~--~3.25 (appropriate for luminosity classes
V~--~I).  Between $T$=30,000~K and $\sim$35,500~K there is a nearly
linear relationship between EW ratio and temperature. The solid line
is a fit to the $\log$($g$)=4.0~--~3.75 models over the range
$T$=27,500~--~40,000~K, and the dashed line is a fit to the models with
$\log$($g$)=3.25~--~3.50 over the range $T$=27,500~--~37,500~K. Above
these maximum temperatures, the linear relation becomes a poor
approximation. Labels near the top of the panel give the corresponding
spectral type according to the observational calibration of
\citet{Martins05} for dwarf and supergiant luminosity classifications.

To obtain the most probable orbital period, we examined each
component's CLEANed radial velocity power spectrum using an
IDL\footnote{The Interactive Data Language (IDL) software is provided
  by Exelis Visual Information Solutions.} program written by
A. W. Fullerton, which makes use of the discrete Fourier transform and
CLEAN deconvolution algorithm of \citet{Roberts1987}. Sufficient
radial velocity and light curve data exist for each system so that
periods are secure and free of aliases.  Spectroscopic solutions were
then computed with the Binary Star Combined Solution Package, {\tt
  BSCSP} \citep{Gudehus01}. The spectroscopic orbital parameters
determined from {\tt BSCSP} were used as initial parameters within the
eclipsing binary package, {\tt PHOEBE} \citep{phoebe}, which is based
on the code of \cite{WD71}, to model the joint light curves and radial
velocity curves. For the light curve parameters, we chose albedos and
gravity darkening coefficients appropriate for radiative atmospheres
(1.0) and limb darkening coefficients interpolated from
\citet{vanHamme93}.  Gravity darkening and albedo were treated as fixed
parameters.  We allowed each system one reflection from the opposite
star and used the Kurucz atmospheric models as the description of the
star's wavelength-dependent emergent flux \citep{Kurucz79}.

We selected the appropriate {\tt PHOEBE} geometrical configuration for
each system based on the shape of the light curve (i.e.,
double-contact for CPR2002~A36 and Schulte~3, and detached for both
MT91~372 and MT91~696). Including only the spectroscopic data for each
system, we used {\tt PHOEBE}'s differential correction method to
refine the spectroscopic solution while allowing period, eccentricity,
separation, epoch of periastron, angle of periastron, systemic
velocity, and mass ratio to vary. From this point, for the detached
systems, we determined the roughly optimized photometric solution by
varying inclination, surface potential, luminosity, and secondary
temperature, using only the photometric data. Schulte~3 and
CPR2002~A36 show evidence of hot spots on the secondary component via
the O'Connell effect in their light curves\footnote{The O'Connell
  effect is defined as the unequal out-of-eclipse maxima
  \citep{oconnell}. Many likely causes have been proposed for the
  phenomenon, but for W~Uma systems the probable cause is a hot spot
  resulting from a mass stream between the two components.},
Struve-Sahade effect\footnote{The Struve-Sahade is defined as the
  variation in a component's spectral profiles as a function of
  phase. The equivalent width of the lines may or may not vary, and
  the effect can affect all lines or a few. Helium absorption lines
  are often affected, and the effect is generally associated with the
  secondary. \citet{Linder2007} and \citet{Bagnuolo1999} provide a
  good summary with examples. } in their \ion{He}{1} lines, and
complex varying H$\alpha$ emission. For the model of these two
systems, we added one spot to the secondary and manually adjusted the
spot temperature, size, and location in an iterative process with the
differential corrections to obtain the roughly optimized
solution. Despite the evidence pointing to the spot being associated
with the secondary (discussed in Sections 4.5 and 4.6), we also
attempted placing the spot on the primary. For both systems, the spot
on the secondary provided the better fit to the data. The spot
parameters were adjusted manually because of the degenerate nature of
the solutions (i.e., spot size, spot temperature, and stellar
effective temperature are degenerate when the analysis is limited to
1~--~2 photometric bandpasses).

Once refined solutions for both the photometric and spectroscopic data
were obtained separately, we optimized the ephemeris by using all data
and varying the period and epoch of periastron. The eccentricity and
angle of periastron were held fixed for MT91~696, CPR2002~A36, and
Schulte~3, but varied for MT91~372.  Finally, we obtained our best-fit
solution by varying the remaining variables that depend on both
spectroscopic and photometric data (i.e., inclination, separation, and
relative luminosities). Owing to uncertainties on the primary
component temperatures, a small grid of solutions was computed for a
range of plausible $T_1$ values.  When a final orbital solution was
obtained, we compared the results with those obtained using {\tt
  Nightfall}, an eclipsing binary package created by Rainer
Wichmann\footnote{ http://www.hs.uni-hamburg.de/DE/Ins/Per \newline
  $\quad$/Wichmann/Nightfall.html} and based on the \citet{WD71} code
assuming typical Roche geometry.  All results from {\tt Nightfall}
were consistent with {\tt PHOEBE} within our final uncertainties. We
obtained formal parameter uncertainties via the instructions in the
{\tt PHOEBE} manual, where fit parameter uncertainties come from the
covariance matrices and derived parameter uncertainties are calculated
via propagation of error. We note that the solutions of Schulte~3 and
CPR2002~A36 suffer degeneracies stemming from probable hot spots and
being limited to 1-2 bandpasses. Further discussion of these
degeneracies are provided in Sections 4.5 and 4.6.

{\tt PHOEBE} solutions provide computed theoretical absolute
bolometric magnitudes. However, because the bolometric magnitudes
represent the components only, rather than the components and
additional spots, we estimated the apparent magnitude for Schulte~3
and CPR2002~A36 from the model by turning off the spot contribution
once a solution was finalized. We then used bolometric corrections
interpolated from the tables of \citet{Martins05} and the {\tt PARSEC}
isochrones \citep{parsec} to determine the absolute visual
magnitudes. The distance to each system may then be derived using the
resulting distance modulus in conjunction with an estimate of the
extinction, $A_V$ (\S4.2).

\subsection{Extinction Calculations}

We estimate the extinction toward each system from the $UBVJHK$
broadband photometry, compiled in Table~\ref{av.tab} from published
sources and from our own $UBV$ photometry, in conjunction with an
adopted spectral energy distribution from \citet{Kurucz92} atmospheres
of the appropriate $T_{eff}$ and $\log$($g$) for its spectral type.
The $JHK$ photometry are drawn from the Two-Micron All Sky Survey
(2MASS) Point Source Catalog\footnote{Because of the low angular
  resolution of 2MASS, there may be flux contributions from unresolved
  components.  However, only in the case of Schulte~3 is there a
  tertiary bright enough and close enough to be considered significant
  (12\%\ of the system light), and even this contribution is
  inconsequential to the reddening results because of the very large
  extinctions.}, while $UBV$ measurements come from various
independent sources. We redden the model atmosphere by either a
\citet[][CCM89]{ccm89} or a \citet[][FM07]{FM07} prescription for
interstellar extinction. \citet{ccm89} provide an analytical
description of the average interstellar extinction law over the
ultraviolet, optical and near-infrared range, parameterized in terms
of two variables, \AV\ and \RV.  \citet{FM07}, by contrast, present a
table of 328 discrete reddening curves toward individual Galactic
sources; they highlight the uniqueness of each sightline, stressing
how generalized curves such as \citet{ccm89} are gross averages over
diverse interstellar dust components. We fit reddened photospheres to
the data for each star using each of these
prescriptions. Table~\ref{av.tab} the best-fit \AV\ and \RV\
values. Our approach is more fully described in \citet{vargas13}, who
find good agreement between these two descriptions of extinction
toward the heavily reddened Galactic cluster Westerlund 2.  They note
how the non-standard value of \RV=3.8 along that sightline results in
a significantly smaller distance than if the canonical \RV=3.1 were
adopted.

Because these targets are variable at the level of 0.5, 0.6, 0.3, and
0.5 mag level for MT91~372, MT91~696, Schulte~3, and CPR2002~A36
respectively, a robust reddening solution ostensibly depends on
acquiring $UBV$ and $JHK$ datasets at the same orbital phase.  The
2MASS $JHK$ observations were made simultaneously at orbital phase,
$\phi$=0.74, 0.84, 0.69, 0.68 for MT91~372, MT91~696, CPR2002~A36, and
Schulte~3, respectively, assuming that the orbital periods have
remained constant since the 1998 June 2MASS observations. The times of
observation are not available in published sources for the majority of
the $UBV$ photometry, except for our new data presented here which we
acquired at orbital phase $\phi$=0.32 and 0.47 for MT91~696 and
Schulte~3, respectively, near JD=2456271.58 early on 2012 December 10
(UT).  Assuming that the orbital periods have remained constant and
that the broadband colors remain unchanged throughout the eclipse
cycle (a plausible assumption given the similarity in temperatures
between components in all cases), we correct our $UBV$ photometry to
the orbital phase of the 2MASS photometry before fitting for the
optimal reddening solution.  These corrected values are listed and
footnoted as such in Table~\ref{av.tab}.  In practice, we find that
the reddening toward all targets is sufficiently large that adjusting
photometry to a common orbital phase is a small correction compared to
the \AV$>$5 mag interstellar contributions to the broadband
magnitudes.  Hence, these corrections make negligible difference on
the final reddening parameters.

The $UBV$ photometry reported in the literature for a given system
shows dispersion beyond the formal uncertainties (0.01~--~0.03 mag
based on published sources), consistent with data being obtained at
different orbital phases and/or real intrinsic variations in the
luminosity of the sources. Furthermore, MT91 note the systematic
differences between their $UBV$ photometry and that of \citet{RLP}.
For purposes of chi-squared minimization, we adopt somewhat generous
uncertainties of 0.05, 0.05, 0.03, 0.01, 0.01, 0.01 mag on the
$UBVJHK$ photometry, respectively.

\begin{deluxetable*}{lrrrr}
\centering
\tabletypesize{\scriptsize}
\tablewidth{0pc}
\tablecaption{$A_V$ and $R_V$ Estimates \label{av.tab}}
\tablehead{
\colhead{Quantity} &
\colhead{MT91~372} &
\colhead{MT91~696} &
\colhead{CPR2002~A36} &
\colhead{Schulte~3}  }  
\startdata            
U                  &  18.01\tablenotemark{a}& 14.44\tablenotemark{a}   & 13.73\tablenotemark{c}   & 12.31\tablenotemark{c}   \\
B                  &  17.14\tablenotemark{a}& 13.97\tablenotemark{a}   & 12.78\tablenotemark{c}   & 11.85\tablenotemark{c}   \\
V                  &  14.97\tablenotemark{a}& 12.32\tablenotemark{a}   & 11.36\tablenotemark{c}   & 10.09\tablenotemark{c}   \\
J                  &  12.51\tablenotemark{b}& 8.53\tablenotemark{b}    & 7.19\tablenotemark{b}    & 6.50\tablenotemark{b}    \\
H                  &   9.92\tablenotemark{b}& 8.14\tablenotemark{b}    & 6.65\tablenotemark{b}    & 6.00\tablenotemark{b}    \\
K                  &   9.60\tablenotemark{b}& 7.89\tablenotemark{b}    & 6.36\tablenotemark{b}    & 5.74\tablenotemark{b}    \\
$A_{\rm V}$(CCM89) &   7.15	      	    & 5.82		       & 6.63			  & 6.16		     \\ 
$R_{\rm V}$(CCM89) &   3.03	      	    & 2.90		       & 3.11			  & 3.11		     \\
$A_{\rm V}$(FM07)  &   7.01	      	    & 5.88		       & 6.56			  & 5.92		     \\ 
$R_{\rm V}$(FM07)  &   2.93	      	    & 3.06		       & 3.31			  & 2.93		     \\
\hline
U                  &	17.19		    & 14.20\tablenotemark{c}   & 13.73\tablenotemark{c}   & 12.30\tablenotemark{e}   \\
B                  &	16.52		    & 13.87\tablenotemark{c}   & 12.81\tablenotemark{d}   & 11.83\tablenotemark{e}   \\
V                  &	15.04		    & 12.36\tablenotemark{c}   & 11.27\tablenotemark{d}   & 10.22\tablenotemark{e}   \\
$A_{\rm V}$(CCM89) &	 7.03		    & 6.30		       & 6.96			& 5.93  		   \\ 
$R_{\rm V}$(CCM89) &	 2.95		    & 3.11		       & 2.92			& 2.92  		   \\
$A_{\rm V}$(FM07)  &	 7.13		    & 5.94		       & 6.46			& 6.07  		   \\ 
$R_{\rm V}$(FM07)  &	 3.81		    & 3.42		       & 3.31			& 3.30  		   \\
\hline
U                  & \nodata		    & 14.00\tablenotemark{f}   & 13.19\tablenotemark{f}   & 12.12\tablenotemark{f}   \\
B                  & \nodata		    & 13.69\tablenotemark{f}   & 12.77\tablenotemark{f}   & 11.77\tablenotemark{f}   \\
V                  & \nodata		    & 12.20\tablenotemark{f}   & 11.15\tablenotemark{f}   & 10.22\tablenotemark{f}   \\
$A_{\rm V}$(CCM89) & \nodata		    & 5.83		       & 6.63			 & 6.13 		    \\ 
$R_{\rm V}$(CCM89) & \nodata		    & 2.98		       & 3.02			 & 3.02 		    \\
$A_{\rm V}$(FM07)  & \nodata		    & 5.75		       & 6.36			 & 6.09 		    \\ 
$R_{\rm V}$(FM07)  & \nodata		    & 3.42		       & 3.64			 & 3.36 		    \\
\hline
Avg. $A_V$ (CCM89) & 7.09$\pm$0.12	    & 5.98$\pm$0.27	       & 6.74$\pm$0.19  	 & 6.07$\pm$0.12	 \\ 
Avg. $R_V$ (CCM89) & 2.99$\pm$0.08	    & 3.00$\pm$0.10	       & 3.09$\pm$0.10  	 & 3.02$\pm$0.09	 \\ 
Avg. $A_V$ (FM07)  & 7.07$\pm$0.12	    & 5.85$\pm$0.09	       & 6.46$\pm$0.23  	 & 6.02$\pm$0.09	 \\ 
Avg. $R_V$ (FM07)  & 3.37$\pm$0.44	    & 3.30$\pm$0.20	       & 3.42$\pm$0.19  	 & 3.19$\pm$0.23	 
\enddata
\tablenotetext{a}{\citet{MT91}}
\tablenotetext{b}{2MASS; data acquired at orbital phase
  $\phi$=0.74, 0.84, 0.69, 0.68 for MT91~372, MT91~696, CPR2002~A36, and Schulte~3,
  respectively, assuming that the orbital periods have remained
  constant since the 1998 2MASS observation.}
\tablenotetext{c}{\citet{RLP}}
\tablenotetext{d}{Tycho-2 \citet{Hog2000}}
\tablenotetext{e}{\citet{Nicolet78}}
\tablenotetext{f}{This work; obtained at orbital phase
  $\phi$=0.32,0.49,0.47 for MT91~696, CPR2002~A36, and Schulte~3,
  respectively We have corrected our measured UBV magnitudes to the
  same orbital phase as the 2MASS JHK measurements using the derived
  orbital solution under the assumption of constant orbital periods
  and constant broadband color.}
\end{deluxetable*} 

Table~\ref{av.tab} lists the photometry and best-fit reddening
parameters \AV\ and \RV\ for each combination of $UBV$ measurements,
paired with the 2MASS $JHK$ photometry, for both the CCM89 and FM07
reddening curves.  Inspection of the individual values reveals
generally good agreement between the reddening parameters computed
using different datasets and the same reddening law.  For example, the
extinction results using CCM89 for MT91~372 are \AV=7.15 and 7.03 mag
for the two available $UBV$ datasets, giving a mean of 7.09. The
results using the FM07 reddening curves are \AV=7.01 and 7.13 mag. The
means of \AV=7.09 and 7.07 mag, respectively, for the CCM89 and the
FM07 curves are nearly identical.  The results using CCM89 for
MT91~696 are \AV=5.82, 6.30, and 5.83 with a mean of 5.98. The results
using the FM07 reddening curves are \AV=5.88, 5.94, and 5.75 with a
similar mean 5.85.  CPR2002~A36 and Schulte~3 have mean reddenings of
\AV=6.74 and 6.07 mag, respectively, using the CCM89 reddening curve
and \AV=6.46 and 6.02 mag, respectively, using the FM07 curves.  We
conclude that there is good agreement between the extinctions derived
using either the MT91 $UBV$ photometry in conjunction with 2MASS $JHK$
and our own photometry with 2MASS. There is also good agreement
between the extinctions derived for MT91~372, MT91~696 and Schulte~3
(dispersions $\sim0.1$ mag) using the two reddening curves, but the
dispersion is somewhat larger for CPR2002~A36, leading to a larger
uncertainty on \AV.

Finally, we note that the best-fit FM07 reddening curves have reduced
chi-squared values in the range 1~--~10 while those for the CCM89
parameterization have much larger reduced chi-squared values of
7~--~50.  We adopt 4 degrees of freedom for the 6 data points in each
case, although this is not strictly correct for FM07 because these
curves are fitted empirical characterizations rather than
parameterizations.  Nevertheless, the FM07 series of curves always
provides a better fit to the data than the average curve of
CCM89.\footnote{The Cygnus region may have an anomalous reddening
  curve \citep{Turner89} so the CCM89 average parameterization could
  be less applicable along this sightline.}  The fitting process
consistently selects curves from the FM07 library corresponding to
early type stars from NGC~6530, suggesting that this sightline
produces reddening similar to that of the dust toward Cyg~OB2. The
mean extinction and reddening of the FM07 fits for MT91~372 are
$7.07\pm0.12$~mag and $3.37\pm0.44$, respectively. For MT91~696, the
FM07 fits yield means of $A_{V}=5.85\pm0.09$~mag and
$R_{V}=3.30\pm0.20$, and this extinction estimate compares favorably
to the $A_{V}=5.84$~mag found by \citet{MT91}, despite their
assumption of $R_{V}=3.0$.  The mean of the FM07 fits for CPR2002~A36
provide $A_{V}=6.46\pm0.23$~mag and $R_{V}=3.42\pm0.19$.  And finally,
for Schulte~3 we find $A_{V}=6.02\pm0.09$~mag and $R_{V}=3.19\pm0.23$,
which is slightly smaller than the $A_{V}=6.24$~mag found by
\citet{TD91}.  \cite{TD91} determined \RV=3.04$\pm$0.09 for dozens of
stars in the Cyg~OB2 vicinity.  Our average value using the CCM89
reddening curve is \RV$\simeq$3.05$\pm$0.05, whereas our average value
using FM07 is \RV$\simeq$3.3$\pm0.1$, somewhat higher but consistent
with values found throughout the region.  Hereafter, we adopt the
best-fit FM07 extinction fits. We note that adoption of the CCM89
reddening law would systematically lower the derived distance modulus
for CPR2002~A36 by about 0.28 mag, but it will have a very small
impact on the distances for MT91~372, MT91~696 and Schulte~3.

\subsection{MT91~372}

\begin{figure}[htpb!]
\centering
\includegraphics[width=\columnwidth,trim=5mm 0 0 0]{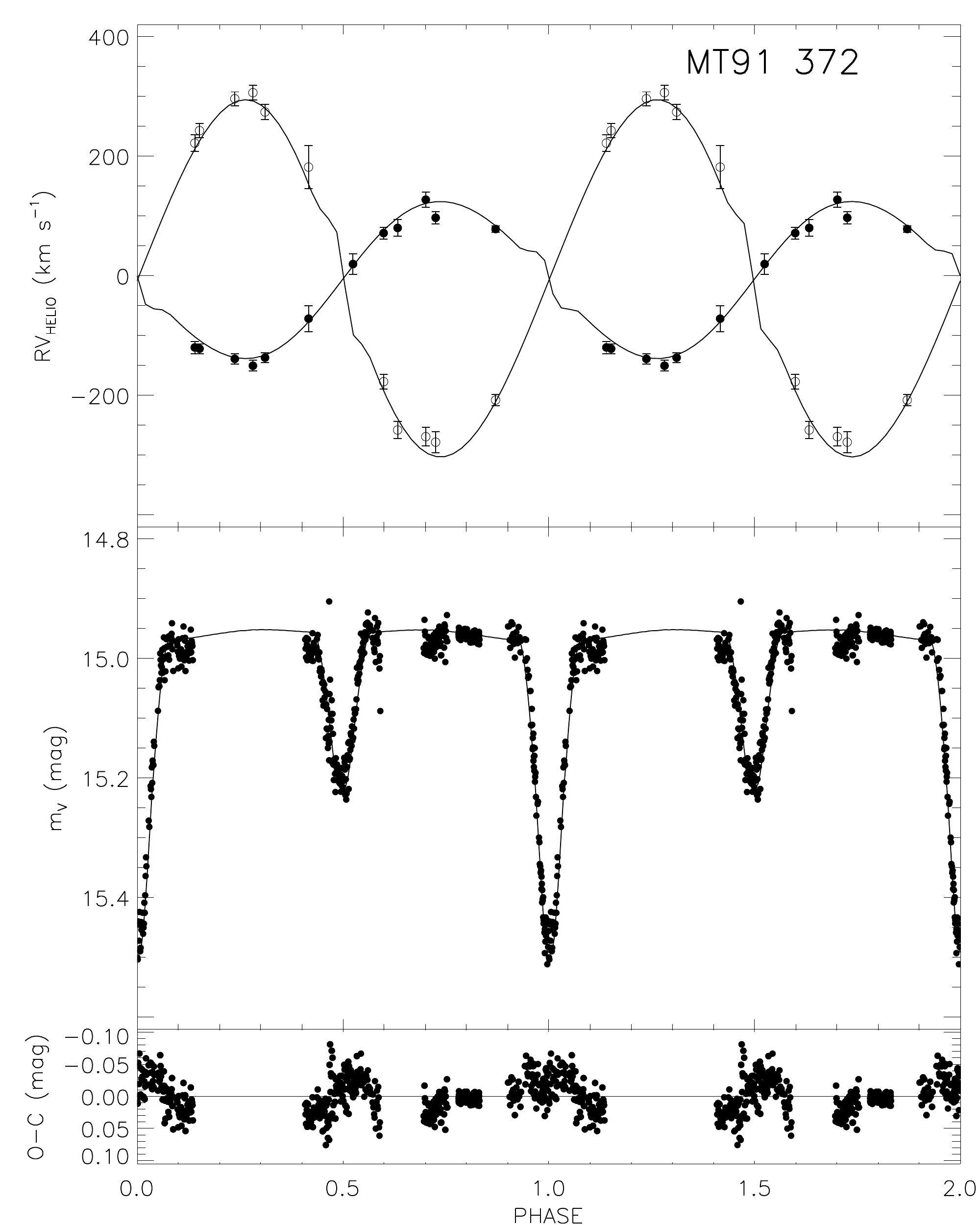}
\caption{Radial velocity curve (top), light curve (middle), and light
  curve residuals (bottom) for MT91~372. Solid and unfilled points
  (top) denote the primary and secondary radial velocities,
  respectively.  Solid curves in each panel show the best-fit {\tt
    PHOEBE} solutions given the adopted parameters, as described in
  the text. The steps at phases of 0.5, 1.0, and 1.5 are a result of
  the Rossiter-McLaughlin
  effect\citep{Rossiter24,McLaughlin24} \label{MT372solution}}
\end{figure}

\begin{figure}[htpb!]
\centering
\includegraphics[width=\columnwidth,trim=10mm 0 0 0]{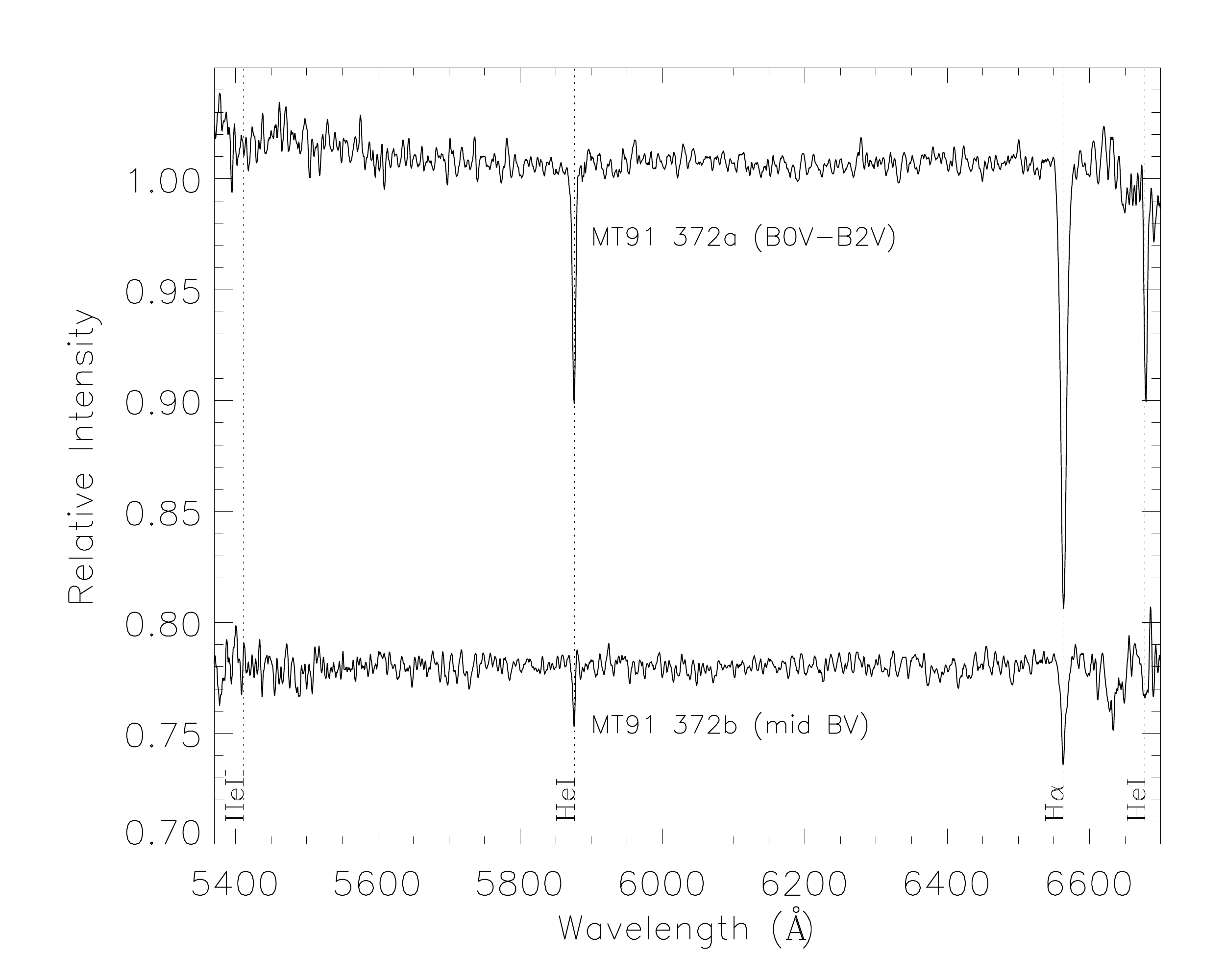}
\caption{Normalized component spectra for MT91~372a and MT91~372b
  using the \citet{GL2006} technique for disentanglement. The spectra
  are shown with arbitrary vertical offsets. \label{MT372component}}
\end{figure}

Of the four eclipsing systems analyzed, MT91~372 is one of the most
ideal for determining distance.  Figure~\ref{MT372solution} shows the
velocity curve data and best-fit model (top), light curve and best-fit
model (middle), and light curve residuals (bottom).  The components
are detached with clear, deep eclipses having $\Delta m\simeq0.2$ for
the secondary eclipse and $\Delta m\simeq0.5$ for the primary eclipse
in V.  Figure~\ref{MT372component} shows disentangled spectra for each
component.  The primary star displays strong \ion{He}{1}~$\lambda$5876
lines relative to H$\alpha$, indicative of a hot star in the range
B0V~--~B2V. While \ion{He}{1} is present in the secondary, this ratio is
smaller, indicating a later spectral type near mid-B.

Because this system is among the faintest and reddest objects in the
Cygnus~OB2 Radial Velocity Survey ($V=14.96$ at its brightest and
\AV$\simeq$7~mag), the SNR of the Survey's WIYN spectra are
insufficient to use the well-calibrated $EW_{4481}/EW_{4471}$ ratios
to measure component temperatures.  We determine a 2$\sigma$ upper
limit of $T_{1}<28,000$~K for the primary based on an upper limit for
the ratio $EW_{5411}/EW_{5876}<0.10$ (calibrated using
Figure~\ref{EWHe}) in a combination of all available WIRO spectra
taken near the same quadrature phase.\footnote{The measured (indicated
  by subscript m) equivalent widths of lines in a composite spectrum
  of two stars, $EW_{m,1}$ and $EW_{m,2}$, yield the true (indicated
  by subscript t) equivalent widths via the ratio of continuum light
  at that wavelength, $l\equiv C_1/C_2$, as $EW_{t,1}=EW_{m,1}(1+l)/l$
  and $EW_{t,2}=EW_{m,2}(1+l)$
  (\citealt[][eqn. 5.98]{Hilditch01}). However, assuming that the
  continuum light ratios of the two very hot stars at 5876\AA\ is the
  same as at 5411\AA, a direct measurement of the ratio
  $EW_{5411}/EW_{5876}$ for each star is essentially independent of
  $l$.}  The available spectra provide no strong lower limits on the
effective temperature of the primary. The presence of
\ion{He}{1}~$\lambda$5876 with an EW of 0.60$\pm$0.02 \AA\ in the
separated primary star spectrum in Figure~\ref{MT372component}
confirms that MT91~372a is a hot star with $T_{1}>18,000$~K based on
the calibration between $T_{eff}$ and \ion{He}{1}~$\lambda$5876
presented in Figure~3 of \citet{Kobulnicky12b}.  The ratio
$EW_{6563}/EW_{5876}\simeq5$ provides similar loose constraints.

In an attempt to provide a more secure lower limit on the effective
temperature of the primary, MT91~372 was observed at WIRO on 2013
August 14 and August 29 for 3 hours ($12\times900$~s) each using a
2000~l~mm$^{-1}$ grating in first order over the wavelength range
6900~--~8000 \AA.  The dates of observation were chosen to cover
portions of the orbital phase (0.09 and 0.85, respectively) where the
temperature-sensitive far-red lines of \ion{He}{1} $\lambda$7065,
\ion{O}{1} $\lambda\lambda$7772,7774,7775 \AA, and \ion{Mg}{2}
$\lambda$7896 fell at locations minimally affected by
atmospheric absorption, interstellar bands, and night sky emission
lines.  No order blocking filter was used as the atmosphere, detector,
and heavy interstellar extinction ($A_{V}=7$~mag) effectively serve to
block wavelengths shorter than 4000 \AA.  Data reduction followed
standard procedures described in Section 2.  Particular care was taken
with the continuum normalization of the spectra to ensure that
line-free portions were used to fit a low-order polynomial to the
spectrum.  Regions of strong atmospheric absorption near
7150~--~7400~\AA\ and 7600~--~7700~\AA\ were excluded from the fit.  From
these spectra we detect \ion{He}{1}~$\lambda$7065 at the expected
level for stars between 20,000~K and 28,000~K, but
\ion{O}{1}~$\lambda\lambda$7772,7774,7775~\AA, and
\ion{Mg}{2}~$\lambda$7896 lines are not detected to upper limits of
$\sim$0.12~\AA.  While this limit does not place useful constraints on
the temperature on the basis of the expected \ion{Mg}{2}~$\lambda$7896
equivalent widths (0.05 \AA\ or less for stars hotter than 20,000~K,
based on Tlusty models), the limits {\it are} useful in the case of
the \ion{O}{1}~$\lambda\lambda$7772,7774,7775 blend which ought to
have a combined EW$>$0.2~\AA\ for stars cooler than 20,000~K.

We conclude that 20,000~K is a secure lower limit on $T_1$.
Accordingly, we adopt a temperature of $24,000\pm3,000$~K for the
primary component, MT91~372a. Given that the system is clearly
detached with a small separation, the components are not likely
filling their Roche lobes. Based on the absence of emission features
we assume that the components are un-evolved.  The joint light curve
and velocity curve analysis to follow confirms that the components
have radii consistent with main-sequence, early B stars.

Radial velocities obtained from WIRO-Longslit spectra between 2008 and
2013 (12 epochs for the primary, 11 epochs for the secondary) produce
power spectra with two strong signals, the likely period at
$\sim$2.218~d and the $1-\nu$ alias at $\sim$1.821~d. The {\tt BSCSP} 
solution also converged on a $2.218$~d period.

Our photometric and spectroscopic data, adopted $T_1$, and starting
orbital parameter guesses were used to compute the best-fit {\tt
  PHOEBE} orbital solution displayed in
Figure~\ref{MT372solution}. The best period for the combined datasets
is $P=2.227490\pm0.000005$~days, and the light curve requires a
non-zero, but low, eccentricity of $e=0.044\pm0.002$.  The best {\tt
  PHOEBE} solution provides a slightly lower mass ratio of
$q=0.44\pm0.01$ (BSCSP yields
$q=0.50\pm0.02$). Table~\ref{allfits.tab} lists the best-fit orbital
parameters, including the orbital period ($P$), the date of periastron
($T_0$), eccentricity ($e$), orbital inclination ($i$), semi-major
axis ($a$), systemic velocity ($\gamma$), and mass ratio
($q$). Table~\ref{allfits.tab} also lists physical parameters for the
system derived from the modeling process, including the secondary
temperature ($T_2$), the radii of both components ($R_1$ and $R_2$),
the masses ($M_1$ and $M_2$), and the bolometric magnitudes
($M_{Bol1}$ and $M_{Bol2}$). Not surprisingly, given the depth of the
eclipses, MT91~372 is nearly edge-on with an inclination of
$i=86\pm1$\deg. This results in computed component masses of
$13.0\pm0.7$~\msun\ and $5.7\pm0.4$~\msun\ and radii of
$4.94\pm0.04$~\rsun\ and $3.34\pm0.06$~\rsun. Coupled with the
estimated primary temperature and the resulting best fit secondary
temperature ($T_{2}=16,600\pm2,300$), the best-fit spectral types are
B1V and B5V \citep{drilling}, agreeing well with our initial
estimates.  The final rows of Table~\ref{allfits.tab} list the adopted
extinction ($A_V$=7.07$\pm$0.12), and the observed out-of-eclipse
apparent magnitude of the system ($m_V=14.96$), the distance modulus
(D.M.), and the resulting distance ($d$) in kpc.  Given the derived
bolometric magnitudes ($-4.87$~mag and $-2.41$~mag for the primary and
secondary, respectively), the distance modulus may be computed in the
standard manner,

\begin{equation}
  \begin{split}
  D.M.& = m_V - M_V -A_V \\
      & = m_V + 2.5~\log(10^{-(M_{Bol1}-BC_{1})/2.5} \\
      & \quad +10^{-(M_{Bol2}-BC_{2})/2.5})- A_V.
  \end{split}
\end{equation}

The resulting distance modulus of $10.62\pm0.28$ equates to a distance
of $1.33\pm0.17$~kpc.  We note here that the uncertainties on
$M_{Bol1}$ and $M_{Bol2}$ are coupled, not independent, since $T_2$ is
derived from $T_1$ via the ratio of eclipse depths. Furthermore, there
is an anti-correlation between bolometric magnitudes (i.e.,
temperatures) and the bolometric corrections (B.C.), meaning that an
error in $M_{Bol}$ is partially canceled by a change in B.C. in the
opposite direction. Consequently, the derived $M_V$ used to
determine the distance modulus is less sensitive than one might
ordinarily assume to the adopted primary effective temperature,
$T_1$. Nevertheless, the relatively large uncertainty on the
temperature of MT91~372A is the dominant source of uncertainty on the
derived distance for this system.

\subsection{MT91~696}

\begin{figure}[htpb!]
\centering
\includegraphics[width=\columnwidth,trim=5mm 0 0 0]{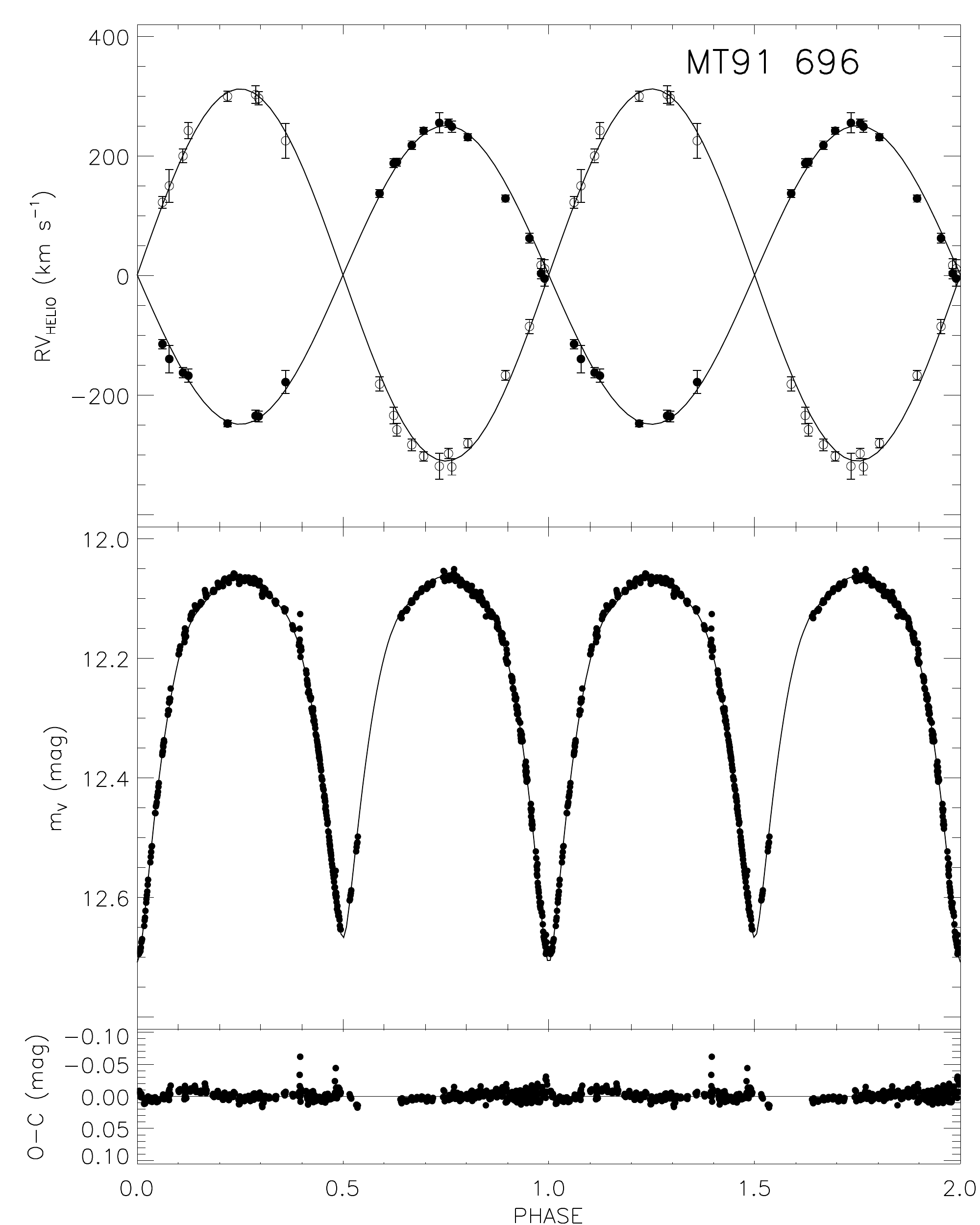}
\caption{Radial velocity curve (top), light curve (middle), and light
  curve residuals (bottom) for MT91~696. Solid and unfilled points
  (top) denote the primary and secondary radial velocities,
  respectively.  Solid curves in each panel show the best-fit {\tt
    PHOEBE} solutions given the adopted parameters, as described in
  the text. \label{MT696solution}}
\end{figure}

When corrected for third light contributions, MT91 696 is also a good
object for determining the distance to Cyg~OB2.  As a relatively
bright ($m_{V}=12.05$ mag), short-period system ($P=1.46918$~d), it
contains components close in spectral type (O9.5V and B0.5V) and
luminosity \citep{Kiminki12a}, the system is nearly edge-on with
eclipse depths of $\sim0.64$~mag.  Figure~\ref{MT696solution} displays
the radial velocity curves (top), light curve (middle), and light
curve residuals (bottom). The solid curves represent the best-fit {\tt
  PHOEBE} model discussed below.  The smooth, continuous variation in
the light curve indicates that the components are near contact.
Velocity and photometric uncertainties are small compared to the
variations.

\begin{deluxetable*}{lllll}
\tabletypesize{\scriptsize} 
\tablewidth{0pc}
\tablecaption{Joint Light and Velocity Curve Analysis for MT91~372,
  MT91~696, CPR2002~A36, and Schulte~3 \label{allfits.tab}}
\tablehead{ 
\colhead{} & 
\colhead{MT91~372} &
\colhead{MT91~696} &
\colhead{CPR2002~A36} &
\colhead{Schulte~3}  } 
\startdata
$P$ (days)             & 2.227582$\pm$0.000005       & 1.469179$\pm$0.000001		  & 4.6749$\pm$0.0006    &   4.74591$\pm$0.00005   \\
$T_O$ (HJD-2,400,000)  & 56100.854$\pm$0.001	     & 54731.6401$\pm$0.0005		  & 55725.8$\pm$0.2	 &   53996.093$\pm$0.006   \\
$e$                    & 0.044$\pm$0.002	     & 0 (fixed)			  & 0 (fixed)    	 &   0 (fixed)    \\
$A$ (R$_\odot$)         & 19.0$\pm$0.3		     & 16.3$\pm$0.2			  & 38.7$\pm$0.8		 &   36.0$\pm$0.9	\\
$q$                    & 0.44$\pm$0.01  	     & 0.80$\pm$0.02			  & 0.62$\pm$0.02	 &   0.42$\pm$0.03   \\
$V_O$ (km s$^{-1}$)     & -7$\pm$2		     & 1.1$\pm$2.7			  & -29$\pm$3		 &   -47$\pm$3   \\
$i$ (degrees)          & 86$\pm$1		     & 86.5$\pm$1.5			  & 73$\pm$1		 &   59.2$\pm$0.3   \\
3rd light              & \nodata		     & 14\%\ (fixed)			  & \nodata		 &   12\%\ (fixed)   \\
$\log(g)$ (P)          & 4.2			     & 4.0				  & 3.4 		 &   3.3   \\
$\log(g)$ (S)          & 4.1			     & 4.0				  & 3.4 		 &   3.3   \\
$T_1$ (K)              & 24,000$\pm$3,000	     & 32,000$\pm$600			  & 30,000$\pm$2,000	 &   38,500   \\
$T_2$ (K)              & 16,600$\pm$2,300            & 30,940$\pm$1,050  	          & 20,600$\pm$600	 &   31,300$\pm$700   \\
$R_1$ (R$_\odot$)       & 4.94$\pm$0.04  	     & 6.36$\pm$0.03			  & 16.4$\pm$0.1	 &   16.5$\pm$0.1   \\
$R_2$ (R$_\odot$)       & 3.34$\pm$0.06  	     & 5.72$\pm$0.02			  & 13.1$\pm$0.1	 &   11.0$\pm$0.1   \\
$M_1$ (M$_\odot$)       & 13.0$\pm$0.7		     & 15.0$\pm$0.2			  & 22$\pm$2		 &   20$\pm$2	\\
$M_2$ (M$_\odot$)       & 5.7$\pm$0.4		     & 12.1$\pm$0.8			  & 14$\pm$1		 &   8.2$\pm$0.9   \\
Spectral Type 1       & B1V                          & O9.5V                              & O9~--~O9.5III:         &   O6V    \\
Spectral Type 2       & B5V                          & B0.5V                              & O9.5~--~B0IV:          &   O9III~--~O9II  \\
$M_{Bol1}$ (mag)        & -4.87$\pm$0.60               & -6.66$\pm$0.14  	          & -8.44$\pm$0.30	 &   -9.54$\pm$0.54   \\    
$M_{Bol2}$ (mag)        & -2.41$\pm$0.68               & -6.29$\pm$0.20  	          & -6.32$\pm$0.30	 &   -7.76$\pm$0.39   \\  
$B.C._1$ (mag)         & -2.38$\pm$0.33               & -3.05$\pm$0.09			  & -2.88$\pm$0.16	 &   -3.59$\pm$0.37   \\
$B.C._2$ (mag)         & -1.45$\pm$0.38               & -2.96$\pm$0.10  	          & -2.01$\pm$0.17	 &   -2.99$\pm$0.15   \\       
$M_{V,1}$ (mag)         & -2.49$\pm$0.27               & -3.61$\pm$0.04  	          & -5.56$\pm$0.15	 &   -5.95$\pm$0.17   \\  
$M_{V,2}$ (mag)         & -0.96$\pm$0.30               & -3.33$\pm$0.10  	          & -4.32$\pm$0.13	 &   -4.77$\pm$0.25   \\  
$M_V$ (mag)            & -2.73$\pm$0.28               & -4.24$\pm$0.07	                  & -5.86$\pm$0.14	 &   -6.26$\pm$0.19   \\  
$A_V$ (mag)            & 7.07$\pm$0.12  	     & 5.85$\pm$0.09			  & 6.46$\pm$0.23	 &   6.02$\pm$0.09   \\
$m_V$ (mag)            & 14.96$\pm$0.02 	     & 12.22$\pm$0.01\tablenotemark{a}  & 11.40$\pm$0.01\tablenotemark{b}	 &   10.36$\pm$0.03\tablenotemark{a,b} \\
$RMS_{V}$ (mag)         & 0.023                       & 0.007                              & 0.015                &   0.013 \\
$RMS_{NSVS}$ (mag)       & \nodata                     & \nodata                            & 0.035                  &   0.043 \\
$RMS_{RV1}$ (km s$^{-1}$) & 10.2                        & 12.9                               & 25.2                 &   32.9  \\
$RMS_{RV2}$ (km s$^{-1}$) & 14.3                        & 19.3                               & 20.5                 &   37.6  \\
D.M. (mag)             & 10.62$\pm$0.28              & 10.61$\pm$0.10			  & 10.80$\pm$0.27	 &   10.60$\pm$0.21   \\   
Dist. (kpc)            & 1.33$\pm$0.17  	     & 1.32$\pm$0.07		          & 1.44$\pm$0.18	 &   1.32$\pm$0.13   \\ 
\enddata
\tablenotetext{a}{component A$+$B only}
\tablenotetext{b}{spot contribution removed}
\end{deluxetable*} 

\begin{figure}
\centering
\includegraphics[width=\columnwidth,trim=10mm 0 0 0 ]{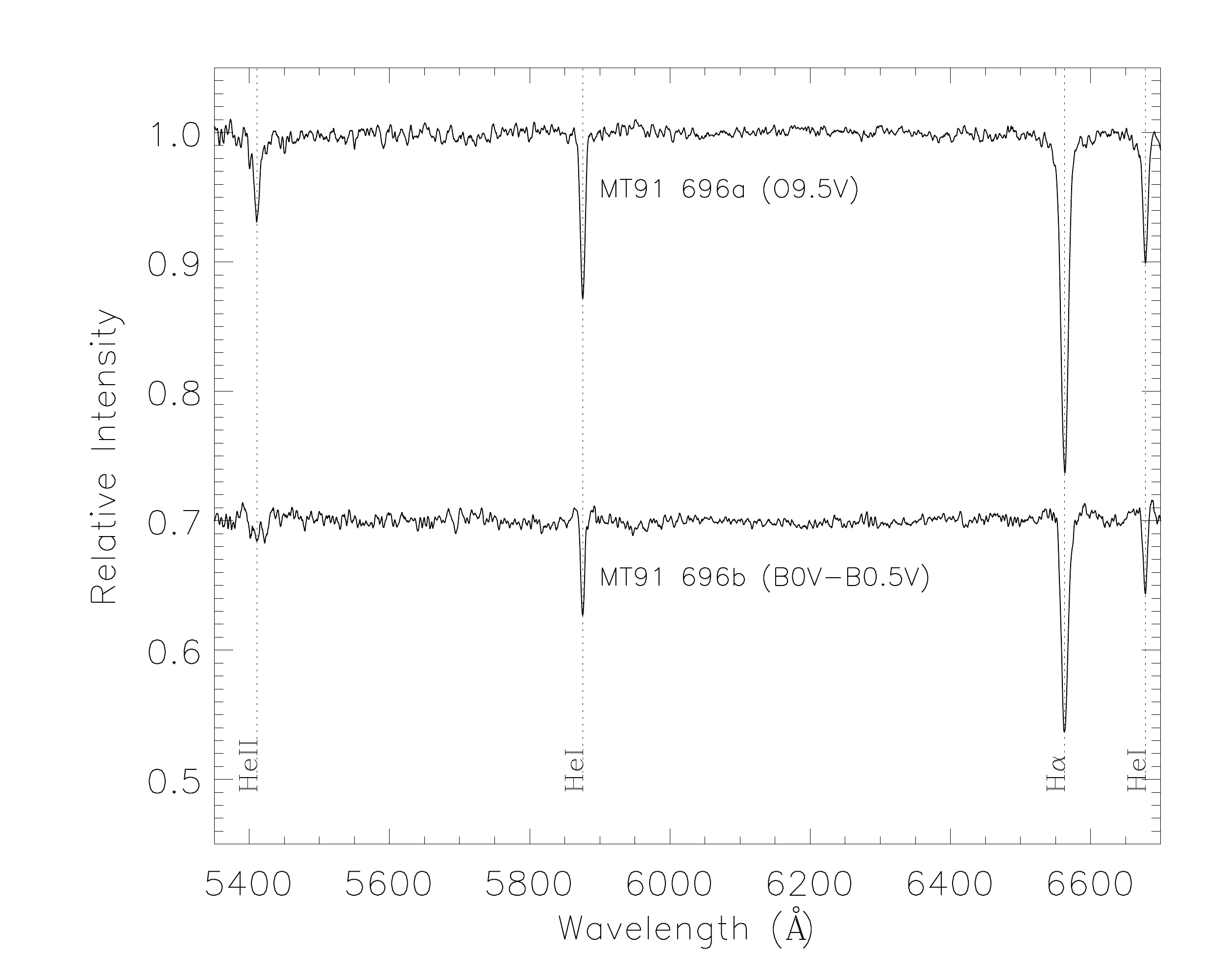}
\caption{Normalized component spectra for MT91~696A and MT91~696B
  using the \citet{GL2006} technique for
  disentanglement.  The secondary spectrum is shown with an arbitrary vertical
  offset for clarity. \label{MT696component}}
\end{figure}

To estimate the component temperatures, we utilized 21 epochs of
WIRO-Longslit spectra taken over nine nights between 2008 August 20
and 2012 October 21 to disentangle the component spectra (shown in
Figure~\ref{MT696component}) and obtain the cross-correlation radial
velocities listed in Table~\ref{specdata.tab}.  In the disentangled
primary spectrum, the EW ratio of temperature-sensitive lines
$EW_{5411}/EW_{5876}$ is $0.45\pm0.05$. We estimate the ratio for the
secondary to be $EW_{5411}/EW_{5876}=0.25\pm0.05$. We also measured
the equivalent width ratio using a composite of four spectra obtained
near a single quadrature phase ($\phi=0.75$). The measured ratios were
higher using this method, $0.76\pm0.07$ and $0.53\pm0.07$ for the
primary and secondary respectively. We adopted the mean of these
ratios, $0.61\pm0.09$ for the primary and $0.44\pm0.09$ for secondary.
Using the relationship shown in Figure~\ref{EWHe}, this yields
$T_{1}=32,000\pm600$~K and $T_{2}=29,800\pm700$~K. The primary
temperature is similar to the 31,884~K implied by the observational
scale of \citet{Martins05} for an O9.5V. Accordingly, we adopt
$T_1=32,000\pm600$ as a fixed parameter in the {\tt PHOEBE} modeling
but leave $T_2$ as a free parameter to be fit.

Before considering the third light contribution from MT91 696C, we
computed the best-fit {\tt PHOEBE} model to the photometry that
includes this extra component.  We then tested various orbital
configurations, and each indicated that the system is just barely
detached. Once the code converged on the best-fit solution, we added
increasingly larger contributions from third light until {\tt PHOEBE}
could no longer converge on a solution and the primary eclipse depth
exceeded 0.75 magnitudes. The highest contribution allowed was 14\% of
maximum out-of-eclipse light (i.e., 14\% of the V=12.051~mag system).
This indicates that the tertiary cannot be brighter than about
V=14.18~mag. Our ephemeris indicates that the $HST$ FGS measurement
was taken at a phase near maximum light. The 0.94$\pm$0.40~mag offset
between the (A+B) binary and the brightest astrometric companion, as
reported by \citet{Caballero2014}, results in a nominal contribution
of 22$^{+20}_{-12}$\% --- consistent with the maximum of 14\% allowed
by the light curve.  We therefore adopt a contribution of 14\% for
component $c$.

Table~\ref{allfits.tab} displays the resulting system parameters for
the best solution to the MT91~696 system after including a constant
14\% third light contribution within the {\tt PHOEBE} model and fixed
zero eccentricity (when allowing the eccentricity to vary, {\tt BSCSP}
and {\tt PHOEBE} both converged on the same zero eccentricity
solution). The final orbital period obtained with all radial velocity
and light curve information is in excellent agreement with the period
reported by \citet{Souza14}, indicating there has been negligible
change in the period over six years. Additionally, while the computed
radii ($R_{1}=6.36\pm0.03$~\rsun\ and $R_{2}=5.72\pm0.02$~\rsun) are
slightly smaller than the theoretical values of $7.3\pm0.1$~\rsun\
(09.5V) and $6.6\pm0.1$~\rsun\ (B0.5V), the computed masses
($M_1=15.0\pm0.2$~\msun\ and $M_2=12.1\pm0.8$~\msun) are in good
agreement with the theoretical values of $16.0\pm0.5$~\msun\ and
$12.9\pm0.4$~\msun\ \citep{Martins05}. Finally, the resulting
secondary temperature, $T_2=30,940\pm1050$~K also agrees well with the
direct $T_2$ measurement from the
\ion{He}{2}~$\lambda$5411/\ion{He}{1}~$\lambda$5876 line ratio
determined above.

The light curve residuals in the lower panel of
Figure~\ref{MT696solution} show small $<$1\% systematic variations
between phase 0 and 0.25, suggesting minor deficiencies in the adopted
model, but additional high-quality photometry will be needed to
further constrain the model parameters. The computed distance modulus
is 10.61$\pm$0.10 mag and corresponds to a distance of
$d$=1.32$\pm$0.07 kpc. For this system, the largest source of
uncertainty on the distance is the uncertainty on the correction for
interstellar extinction.

\subsection{CPR2002 A36}

Ideally, eclipsing binary distances would make use of detached systems
which have simple geometries, i.e., the stars are spherical and the
light curves are unaffected by complications involving mass loss, hot
spots produced by radiation or mass transfer streams, and variations
in the surface potentials of stars approaching or undergoing Roche
lobe overflow.  Unfortunately, four of the six known eclipsing SB2
systems in the direction of Cyg~OB2 contain evolved components and are
probable contact or over-contact systems.  Therefore, we proceed to
model the CPR2002~A36 and Schulte~3 eclipsing SB2 systems while
cautioning that the interpretation of the velocity curves and light
curves is subject to larger uncertainties.

\begin{figure}[htpb!]
\centering
\includegraphics[width=\columnwidth,trim=5mm 0 0 0]{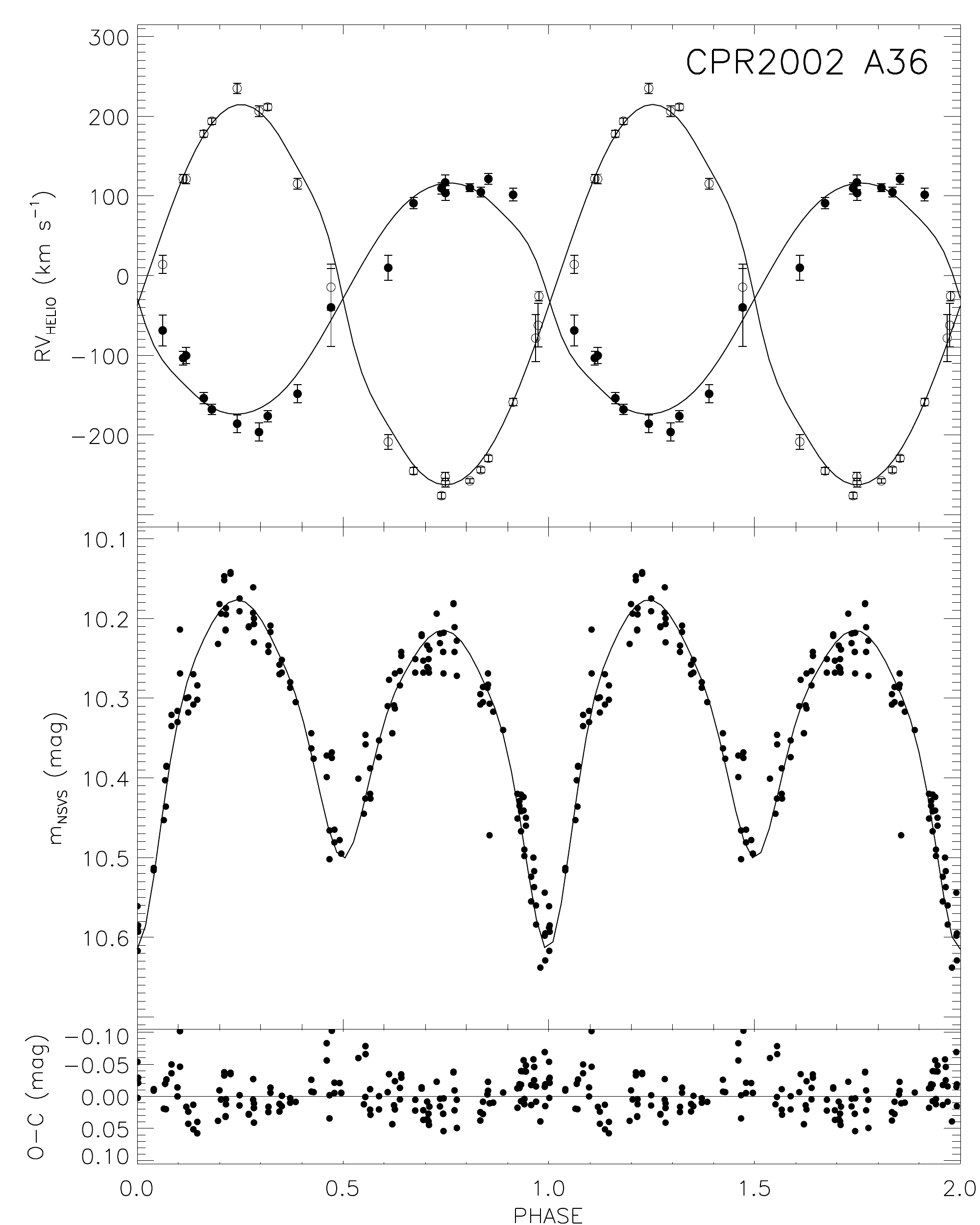}
\caption{Radial velocity curve (top), light curve (middle), and light
  curve residuals (bottom) for CPR2002~A36. Solid and unfilled points
  (top) denote the primary and secondary radial velocities,
  respectively.  Solid curves in each panel show the
  best-fit {\tt PHOEBE} solutions given the adopted parameters, as
  described in the text. \label{A36solution}}
\end{figure}

Figure~\ref{A36solution} shows the best-fit light curve data and model
(top), velocity curve data and model (middle), and light curve
residuals (bottom) for CPR2002~A36. The light curve is characteristic
of an over-contact system with a difference in eclipse depths that
indicates a significant difference in component effective
temperatures. Additionally, we see an O'Connell effect that results in
an $\sim$0.04~mag difference in maxima. There is also considerable
dispersion in the $NSVS$ magnitude at nearly all phases, exceeding the
nominal 0.01 mag uncertainties and suggestive of a high level of
intrinsic temporal variability. Finally, because of the over-contact
nature of the system, CPR2002~A36 shows strong, varying H$\alpha$
emission and a Struve-Sahade effect in the \ion{He}{1} lines.

\begin{figure}[htpb!]
\centering
\includegraphics[width=\columnwidth]{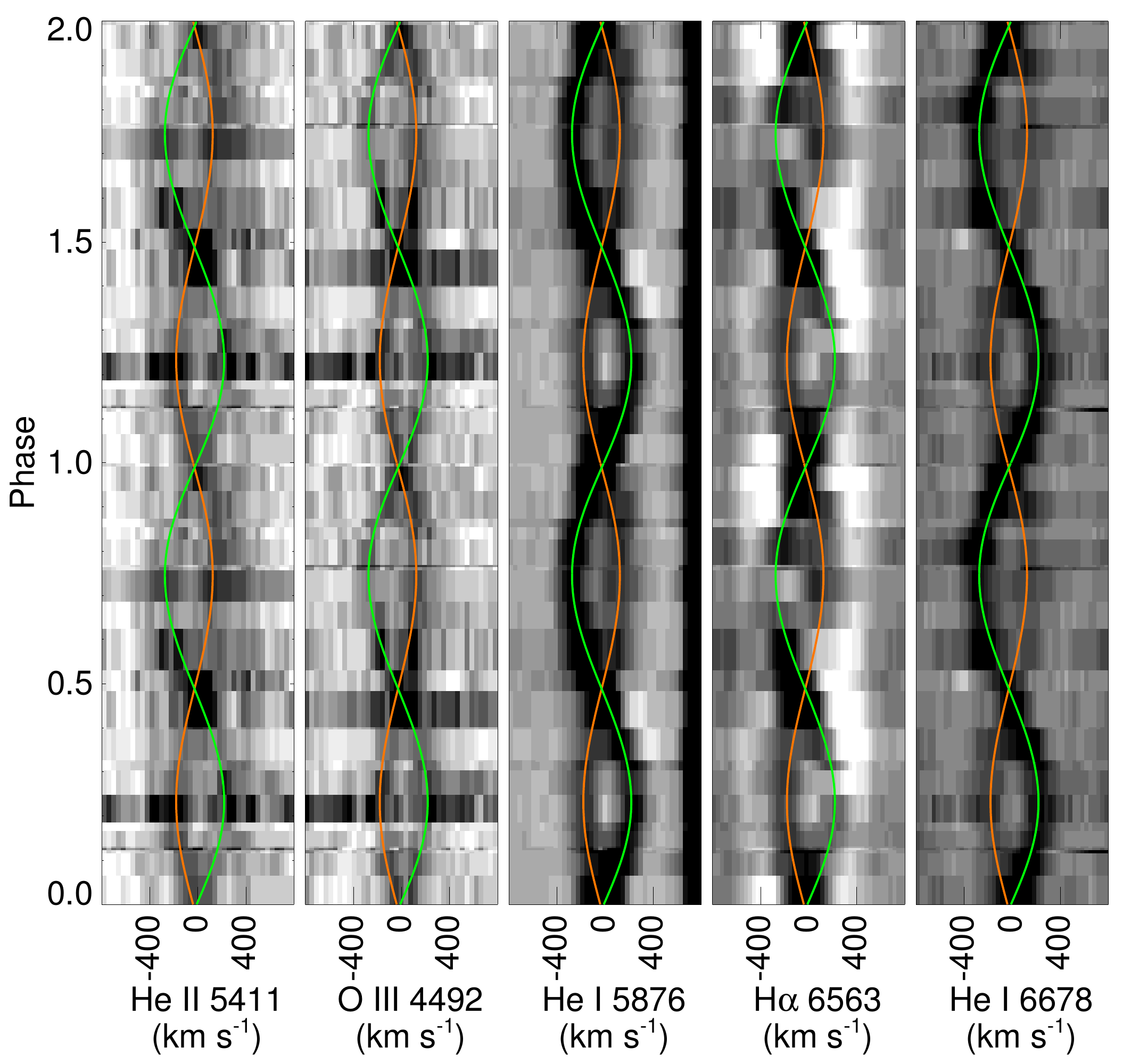}
\caption{A sequence of spectra for CPR2002~A36 ordered by orbital phase,
  showing five spectral lines with the spectroscopic orbital solution
  overplotted in orange for the primary and green for the secondary. 
  \label{A36PhaseSeries}}
\end{figure}

The emission and Struve-Sahade effect are mostly responsible for the
disagreement between the observed and calculated radial
velocities. Emission partially fills the spectral lines and skews the
\ion{He}{1} line centers and is most significant for phases
$\sim$0.8~--~1.1 when the secondary is heading into eclipse. This
emission is easiest to see at H$\alpha$, shown in the panel second
from right in Figure~\ref{A36PhaseSeries} (colored white in this
inverse greyscale image), a phased spectral sequence using all of the
WIRO-Longslit spectra folded with the final solution period. The
figure also shows four additional spectral lines in the
$\lambda\lambda$~5400~--~5700~\AA\ range: \ion{He}{2}~$\lambda$5411,
\ion{O}{3}~$\lambda$5592, \ion{He}{1}~$\lambda$5876, and H$\alpha$,
and \ion{He}{1}~$\lambda$6678.  The orange and green curves represent
the final primary and secondary solutions, respectively. From a phase
of $\sim$0.25~--~0.6, a singular emission feature is evident and
appears to be associated with the redshifted secondary. At phase
0.85~--~1.1, there appear to be multiple emission components with the
dominant one still associated with the secondary. At
\ion{He}{1}~$\lambda$5876 (middle panel), the emission results in the
best Gaussian fit being centered at a longer wavelength than the true
center. Because the primary component's \ion{He}{1} absorption is weaker
and shifted less, the \ion{He}{1} profiles are more distorted at these
phases, and therefore, computed radial velocities are more
skewed. The emission, Struve-Sahade effect, and O'Connell effect may
be adequately explained by a hot spot or a stream of optically thin
ionized gas near or trailing the secondary.

An advantage to creating a phased spectral image is that weak lines
that would normally be difficult to identify amongst the continuum
noise and various interstellar/DIB lines are more easily identified
and associated with the correct component. For instance,
\ion{O}{3}~$\lambda$5592 is difficult to identify with certainty in
many of CPR2002~A36's spectra, but its presence is easily revealed in
both components in Figure~\ref{A36PhaseSeries}. The full phased
spectral image also reveals moderately weak
\ion{C}{4}~$\lambda\lambda$5801, 5812 absorption and very weak
\ion{C}{3}~$\lambda$5696 emission in the primary, the presence and
strength of which, with \ion{He}{2}~$\lambda$5411, indicate the
primary may be slightly hotter than the B0 estimated in
\citet{Kiminki09}, such as O9~--~O9.5 \citep{Walborn80}. Given the
strength of the metal lines in these spectra and of the \ion{Si}{4}
and \ion{C}{3} lines shown in Figure~4 of \citet{Kiminki09}, we revise
our estimate of the primary's luminosity class to III and adopt a
full spectral classification of O9~--~O9.5III. For the secondary, the
full phased spectral image revealed the presence of weak
\ion{N}{2}~$\lambda\lambda$5667, 5670, 5680. Given these, the weak
\ion{He}{2}~$\lambda$5411, and the very weak \ion{O}{3}~$\lambda$5592,
we revise and adopt the spectral type of O9.5~--~B0IV: for the
secondary.

We estimate the effective temperatures for this system in a similar
manner to MT91~696 via the ratio of helium equivalent widths. We
measured equivalent widths at both quadrature phases, obtaining
$EW_{5411}/EW_{5876}=0.52\pm0.09$ and
$EW_{5411}/EW_{5876}=0.48\pm0.04$ for the primary and secondary,
respectively. As both components of CPR2002~A36 have been classified
as likely evolved, we use the dashed line in Figure~\ref{EWHe} to
estimate the temperatures as $T_1$=29,400~K and $T_2$=29,200~K with
uncertainties of approximately 500~K. These estimates are in rough
agreement with the theoretical temperatures for an O9III
(31,000$\pm$800~K) and O9.5IV (30,800$\pm$500), but disagree with the
temperature difference indicated by the difference in eclipse
depths. However, considering the lower signal-to-noise of
\ion{He}{2}~$\lambda$5411 in many of the spectra and the fact that the
Struve-Sahade effect raises the inherent uncertainty in equivalent
width measurements of \ion{He}{1}, and given the theoretical
temperatures \citep{Martins05} for the estimated spectral types above,
we adopt $30,000\pm2,000$~K for $T_{1}$ and estimate
$29,000\pm2,000$~K for $T_{2}$.

Our nominal {\tt PHOEBE} model is a double-contact configuration with
a hot spot on the trailing side of the secondary (130\deg\ from the
point of contact on the equator). In this scenario, the visibility of
the spot coincides well with the presence of emission in
Figure~\ref{A36PhaseSeries}. The spot may indicate the transfer of
mass from the primary. The final solution is fit with a spectroscopic
period of $P=4.6749\pm0.0006$~d and eccentricity fixed at zero. We
were unable to reconcile a phase shift of 0.157 (17.6~hr) between the
NSVS photometry and our radial velocity data with {\tt
  PHOEBE}. Therefore, we adopted the ephemeris obtained from {\tt
  BSCSP} and applied an arbitrary 0.1565 phase shift to the NSVS
photometric data. Similarly, we applied an arbitrary 0.0506 phase
shift to the RBO $V$-band photometry as well. It is possible that the
system has a changing period that we have not accounted for, but its
impact is negligible on the results presented here. Additionally, we
find that the $V$-band photometry and the radial velocity curves are
better fit with a slightly higher inclination than the NSVS photometry
($i=72.9$\deg\ versus $i=69.6$\deg). Because NSVS photometry is
filterless and only approximate to Johnson-R band \citep{Wozniak2004},
we adopt the solution with the higher inclination. This also has
negligible impact on the results presented here. In
Figure~\ref{A36solution}, we show the NSVS photometry instead of the
$V$-band photometry because its full phase coverage better illustrates
the solution's overall agreement with the observed light curves.

The final solution yields a secondary temperature of
$T_2$=20,600$\pm$600~K, contrasting significantly with the
$29,000\pm2,000$~K estimated by helium equivalent width ratios, but
relative to $T_{1}$, it is in better agreement with the temperature
difference indicated from the eclipse depths. No combination of spot
parameters and overall model allowed for a higher secondary
temperature, but we admit that fitting the spot manually limits the
parameter space explored.  The computed masses $M_1$=$22\pm2$~\msun\
and $M_2$=$14\pm1$~\msun\ are in good agreement with the expected
values from \citet{Martins05}. However, the radii,
$R_1$=$16.4\pm0.1$~\rsun\ and $R_2$=$13.1\pm0.1$~\rsun, are larger
than the expected values and more characteristic of O9II and O9.5III.
However, given the uncertainties on the spectral types, and the fact
that each component is filling its Roche lobe, these values are not
entirely surprising. Using the computed bolometric magnitudes in
Table~\ref{allfits.tab}, we compute a distance for this system of
$1.44\pm0.18$ kpc, where the uncertainty on distance is dominated by
the relatively large uncertainty on interstellar extinction
($\sigma_{A_V}$=0.23 mag) and the uncertainty on primary temperature
($\sigma_T=$2000~K).  Had we adopted the CCM89 reddening law for CPR
A36 instead of FM07, the resulting distance would drop to 1.27 kpc. 

\subsection{Schulte~3}

\begin{figure}[htpb!]
\centering
\includegraphics[width=\columnwidth, trim=5mm 0 0 0 ]{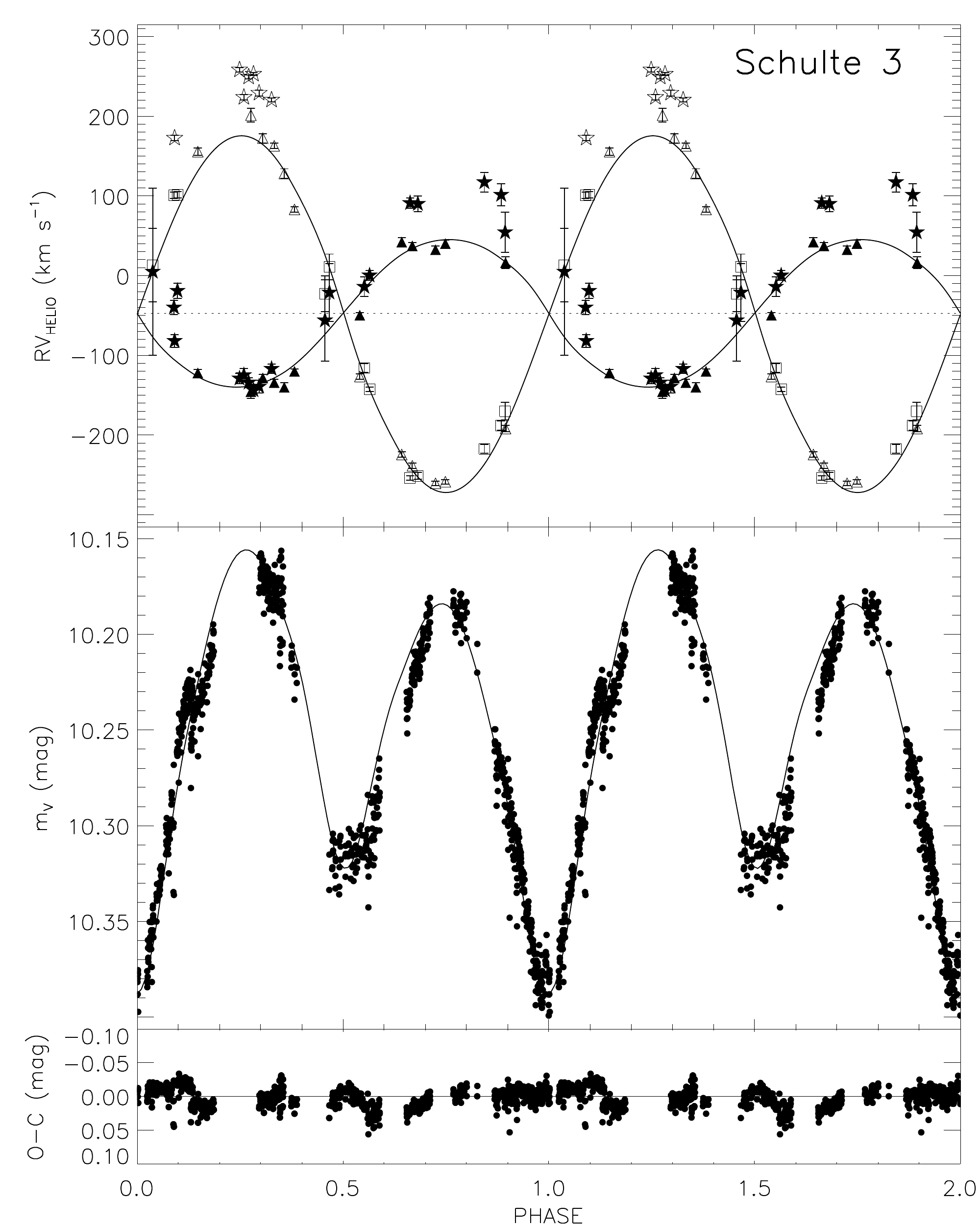}
\caption{Radial velocity curve (top), light curve (middle), and light
  curve residuals (bottom) for Schulte~3. Solid and unfilled points
  (top) denote the primary and secondary radial velocities,
  respectively. Squares represent observations with
  WIRO-Longslit, triangles represent observations taken with
  WIYN-Hydra, and stars represent the observations excluded from
  fitting. Solid curves in each panel show the
  best-fit {\tt PHOEBE} solutions given the adopted parameters, as
  described in the text. \label{S3solution}}
\end{figure}

\begin{figure}[htpb!]
\centering
\includegraphics[width=\columnwidth]{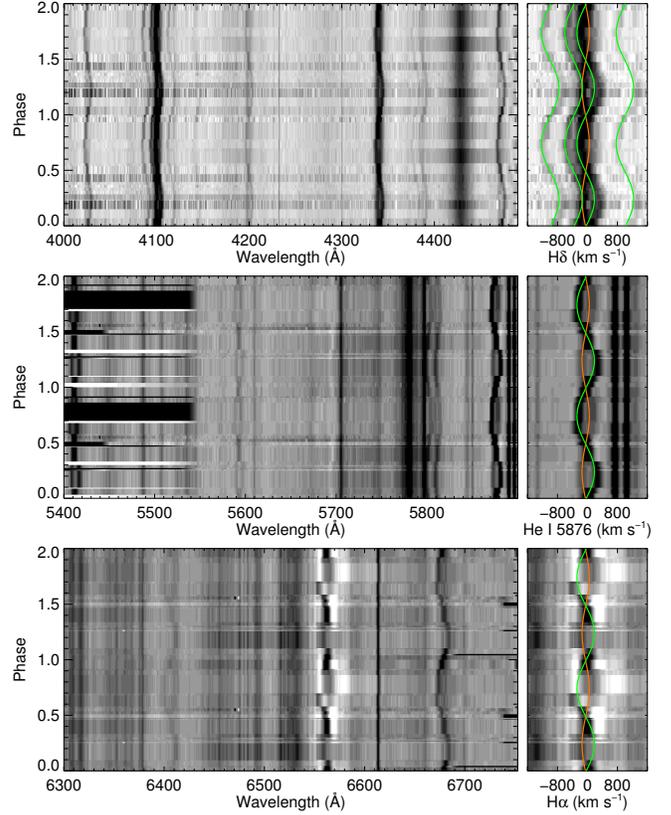}
\caption{A sequence of spectra for Schulte~3 ordered by orbital phase,
  showing nearly the entire wavelength range covered by our WIYN and
  WIRO observations. Right panels illustrate the quality of the
  overall solution with spectroscopic orbital solution overplotted in
  orange for the primary and green for the
  secondary. \label{S3timeseries}}
\end{figure}

Schulte~3 shows many similarities to CPR2002~A36, including highly
variable H$\alpha$ emission, a Struve-Sahade effect, an O'Connell
effect, and a similar period
($4.74591\pm0.00005$~d). Figure~\ref{S3solution} shows the velocity
curve data and model (top), light curve data and model (middle), and
the light curve residuals (bottom). In the top panel, squares
represent WIRO data, triangles represent WIYN data, and stars
represent WIRO data excluded when determining the {\tt BSCSP} and {\tt
  PHOEBE} solutions. We excluded these data because of the distortion
of the \ion{He}{1}~$\lambda$5876 profiles near quadrature. Like
CPR2002~A36, this distortion, the Struve-Sahade effect, more strongly
affects the primary radial velocities and is more easily seen in
H$\alpha$, shown in the bottom panels of Figure~\ref{S3timeseries}
(colored white in this inverse greyscale image). The figure is a
phased spectral sequence using all of the Hydra and WIRO-Longslit
spectra folded with the final solution period, where the orange and
green lines represent the primary and secondary spectroscopic
solutions plotted over various spectral lines, respectively. The
emission in H$\alpha$ and \ion{He}{1} appears at slightly different
phases than with CPR2002-A36, indicating a different spot geometry and
location. In the Figure, H$\alpha$ shows a singular emission feature,
which appears to be associated with the redshifted secondary, at
phases $\sim$0.25~--~0.5. At phases 0.6~--~1.0, it shows multiple
emission components with the dominant one still associated with the
secondary. This emission is also present in \ion{He}{1}~$\lambda$5876,
but it is more difficult to see in the Figure owing to its weaker
presence in the wings around this line. Regardless, the primary
component's \ion{He}{1} radial velocities are skewed at these phases
for the same reasons given for CPR2002~A36. We obtained 20 epochs of
WIYN/Hydra spectra and 24 epochs of WIRO-Longslit spectra. Radial
velocities were measured for 12 and 19 of these spectra, respectively.
However, because of the distorted line profiles we utilized only WIYN
radial velocities for the primary and excluded all secondary
\ion{He}{1}~$\lambda$5876 radial velocities near $\phi=0.25$. The
quality of the final solution is demonstrated with the good agreement
between the radial velocity curves and the spectroscopic solutions in
Figure~\ref{S3timeseries}. In addition to H$\alpha$ in the
bottom-right panel, the top-right panel shows \ion{Si}{4}~4089~\AA,
\ion{N}{3}~4097~\AA, H$\delta$~4101~\AA, and \ion{Si}{4}~4116~\AA, and
the middle right panel shows \ion{He}{1}~5876~\AA. We provide all
radial velocities in Table~\ref{specdata.tab}, with a note to
designate the observations excluded from fitting. The table also
includes the observed minus calculated values of the excluded points
for reference.

\begin{figure}
\centering
\includegraphics[width=\columnwidth,trim = 10mm 0 0 0 ]{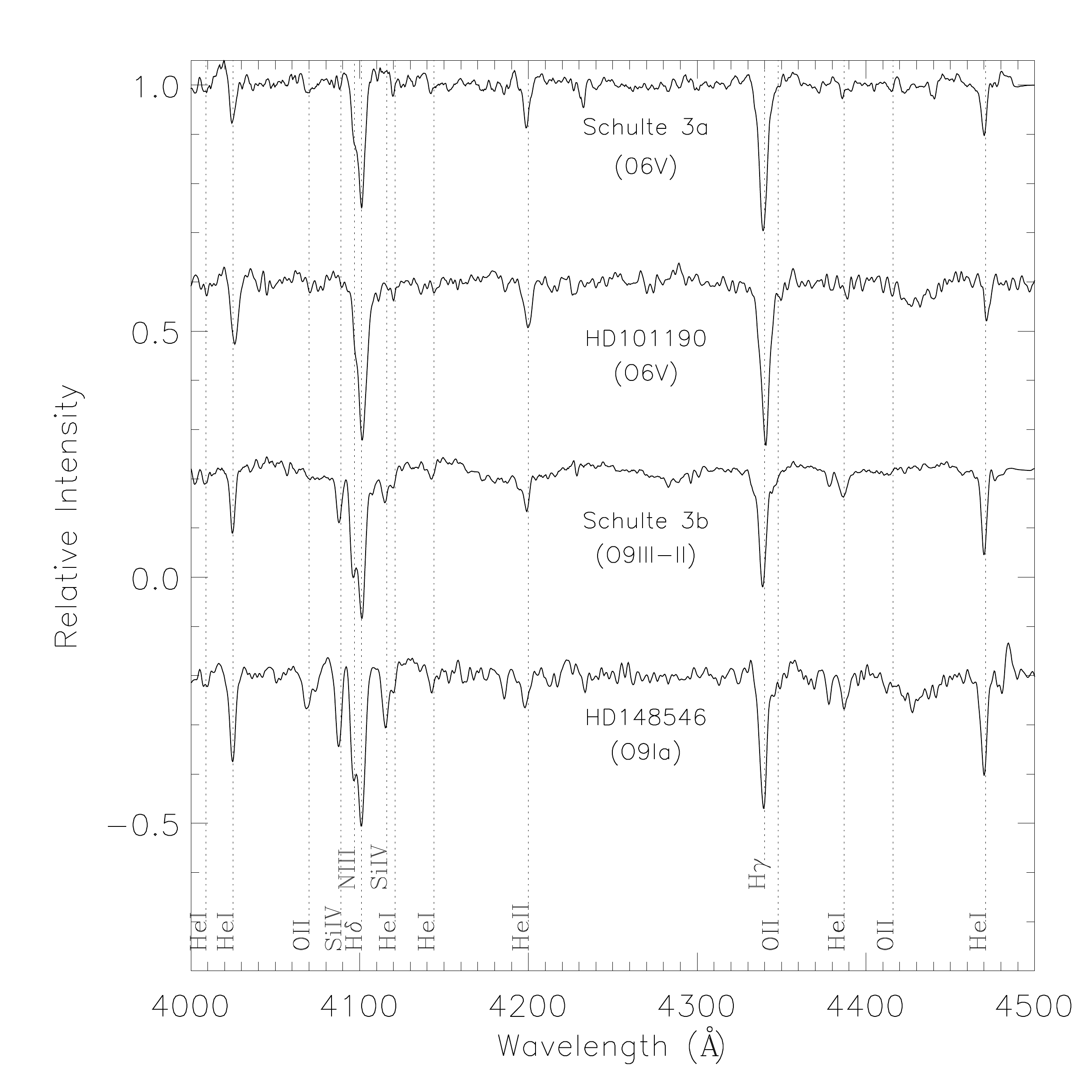}
\caption{Normalized component spectra for Schulte~3A and Schulte~3B using the
  \citet{GL2006} technique for disentanglement. Comparison standard
  spectra from the atlas of \citet{Jacoby84} appear below each
  component spectrum. The spectra are shown with arbitrary vertical
  offsets. \label{S3component}}
\end{figure}

We also utilized the WIYN spectra, after removing contributions from
interstellar \ion{Ca}{2} and the diffuse interstellar band (DIB) at
$\lambda$~4428~\AA, to disentangle the component spectra. Because the
variability of the \ion{He}{1} line depths more strongly affects the
primary than the secondary, \ion{He}{1} is underrepresented in the
disentangled primary spectrum. The spectrum shows
$EW_{4200}/EW_{4026}>1$, indicating a temperature class of early to
mid-O. However, taking into account that \ion{He}{1} is
underrepresented, the ratio may be closer to unity, indicating a
temperature class near O6~--~O7. Additionally, the weak \ion{N}{3}
with absent \ion{O}{2}, \ion{N}{2}, and \ion{Si}{4}, indicates a
likely luminosity class of V. The disentangled secondary spectrum
shows $EW_{4200}/EW_{4144}$ near unity, indicating a temperature class
near O9, and the strength of \ion{Si}{4} and \ion{N}{3} lines relative
to the neighboring \ion{He}{1} lines indicates an evolved luminosity
class, likely III~--~II. The disentangled component spectra are shown
at top and third from top in Figure~\ref{S3component}. The spectra
show good agreement with the spectral standards of HD101190 (O6V;
second from top) and HD148546 (O9Ia; bottom) from the \citet{WF90}
digital atlas. We note that HD101190 has a luminosity of Ia rather
than III, resulting in stronger metal lines, but it is the closest
standard in the library.  Interestingly, the secondary luminosity
class suggests that the present secondary may have been the more
massive component at one point.

As with the other systems, we measured the effective temperatures
using the ratio of helium line equivalent widths in spectra obtained
near quadrature. We find that $EW_{5411}/EW_{5876}$ is 1.06$\pm$0.06
for the primary and 0.95$\pm$0.04 for the secondary. Here, the
Struve-Sahade effect either does not impact our equivalent width
measurements or it impacts them equally, as the variation in
\ion{He}{1} line profile associated with the primary appears to be
limited to line depth and width changes only; the overall equivalent
width does not change between quadrature phases. With these equivalent
width ratios and the luminosity class estimates, the relations
provided in Figure~\ref{EWHe} yield temperatures of
$T_{1}=34,500\pm500$~K and $T_{2}=32,000\pm500$~K for the primary and
secondary, respectively. The secondary estimate is nearly in agreement
with the theoretical value of $31,300\pm600$~K from
\citet{Martins05}. However, the primary estimate is lower than the
theoretical value for an O6V by approximately 4,000~K, which makes the
temperature more appropriate for a slightly later type of O7
\citep{Martins05}. Given that we cannot fully quantify the effect that
emission plays with the equivalent width ratios, we also consider that
the effective temperature may be as high as 38,500~K for the
primary. Therefore, we computed a grid of solutions for primary
temperatures 34,000~K~--~38,500~K. In these solutions, the secondary
temperatures from $\sim$28,000~K~--~35,500~K depending on spot
parameters and model used.

The light curve for Schulte~3 is unambiguously characteristic of a
contact/overcontact system.  It exhibits relatively small eclipse
depths of $\sim$0.23~mag and $\sim$0.17~mag in the raw light curve,
suggesting a moderate inclination. Similarly to CPR2002~A36, the light
curve also displays an O'Connell effect ($\sim$0.02~--~0.03~mag) with a
brighter primary maximum and exhibits real photometric variability in
excess of the nominal uncertainties. Photometry obtained at similar
orbital phase across different orbits exhibit offsets of up to 0.02
mag, signifying intrinsic variability. This is probably a triple
system, and therefore the light curve (and to some minor extent, the
velocity curves) is further complicated by the presence of Schulte~3C
which contributes $\simeq$12\% to the $V$-band light. We used {\tt
  PHOEBE} to account for the third light contribution.

We explored various combinations of temperatures for the primary,
secondary, and spot, various spot configurations, and both W-Uma and
double-contact configuration models for Schulte~3. Because the
emission features appear to be associated mainly with the secondary,
we placed the relatively large (49~--~60\deg) hot spot on that
component, with a center between 24\deg\ and 45\deg\ from the northern
pole and between 20\deg\ and 35\deg\ counter-clockwise from the point
of contact. The adopted location of this hotspot coincides well with
the presence of emission seen in H$\alpha$. In the end, we adopted the
solution with $T_{1}=38,500$~K and $T_{2}=31,300$~K from
\citet{Martins05} because our confidence in the spectral types
exceeded that of the compromised equivalent width
ratios. Additionally, the spot placement for this solution is nearer
to the point of contact (45\deg\ from the northern pole and centered
35\deg\ in the direction of rotation) than the other models, which
makes sense if the primary is heating the secondary. This solution
also provides the most consistent overall picture. The spot is similar
in temperature to the primary ($T_{spot}=37,560$~K), just as we would
expect if it originates from heating, and both the secondary and
combined secondary+spot temperatures ($T_{2+spot}=32,615$~K) agree
with the value obtained from the less compromised secondary equivalent
width ratios. Finally, the combined secondary+spot and primary
temperatures are in better agreement with the temperature difference
implied by the light curves. This solution yields a distance of
1.32~kpc.

However, if we accept the temperatures estimated strictly from
equivalent widths, (i.e., $T_{1}=34,500$~K and $T_{2}=32,100$~K), then
our best fit is a double contact configuration with a spot inclined
$\sim$23\deg\ from the northern pole, centered $\sim$20\deg\ in the
direction of rotation from the point of contact, and with a radius of
$\sim$49\deg. The average secondary+spot temperature is
$\sim$34,800~K. This does not agree with the significant difference in
component temperatures suggested by the light curve. Additionally, the
spot temperature required with this geometry is nearly 45,000~K, which
may be possible with a mass stream origin but not heating. The computed
distance for this scenario is 1.27~kpc.

The computed distance to all explored configurations was between
1.21~kpc and 1.36~kpc, with the closest distance resulting from a
solution with $T_{1}=34,000$~K, $T_{2}=28,596$~K, and
$T_{spot}=34,315$~K. In all solutions, most light-curve-derived
orbital parameters had little variation (e.g., $i\simeq59$\deg,
$R_{1}\simeq16.5$~\rsun, and $R_{2}\simeq11$~\rsun). The largest source
of variation/uncertainty originates from the temperatures and
resultant bolometric magnitudes, with the second largest source being
the high uncertainty in interstellar reddening. Because of the
degeneracies present in this system, when computing uncertainties, we
folded in the full variation of all values for all explored
configurations. We provide these with the full list of orbital
parameters in Table~\ref{allfits.tab} and adopt a distance of
$1.32\pm0.13$~kpc to Schulte~3. Additionally, the light curve
residuals are systematic beyond the stated uncertainties, indicating
additional physical effects that are not correctly reproduced in any
of the explored models.

\section{Discussion}

The weighted mean distance of the four eclipsing binaries analyzed
here is 1.33$\pm$0.06 kpc.  The unweighted mean distance is 1.35 kpc.
This value is broadly consistent with the 1.3~--~1.5 kpc distances
measured for radio masers in pre-main-sequence objects in the
surrounding molecular clouds \citep{Rygl12}.  It is also in good
agreement with the \citet{Dzib12} estimate of 1.3~--~1.4 kpc toward the
massive binary Schulte~5. However, our mean distance is somewhat
smaller than many measurements in the literature which lie in the
range 1.4~--~1.7 kpc.  In particular, \citet{Hanson03} obtained a
spectrophotometric distance of 1.45 kpc toward an ensemble of Cyg~OB2
stars.  Interestingly, \citet{Hanson03} noted that had she adopted the
(at that time, newly revised) cooler stellar effective temperature
scale of \citet{Martins02, Martins05} this would place Cyg~OB2 at 1.2
kpc (D.M.=10.44).  She rejected this possibility, concluding that such
a small distance would reduce the luminosity of Cyg~OB2 O supergiants
that have known mass loss rates and create a problem for the
theoretical understanding of the mechanisms that drive massive star
winds.  However, given recent downward revisions of the stellar mass
loss rates \citep{Puls2008} it is no longer clear that this is a
compelling reason for rejecting a smaller distance.  If we were to
adopt the \citet{Hanson03} distance of 1.20 kpc and correct it upward
by a factor of 1.08 to account for the additional luminosity of
secondary stars in close massive systems as recommended by
\cite{Kiminki12b}, the resulting distance is 1.30 kpc, in excellent
agreement with our current eclipsing binary estimate. 

We note here that \citet{Wright15} derived slightly lower
extinctions of $A_{V}=6.98^{+0.50}_{-0.26}$~mag and
$5.81^{+0.32}_{-0.29}$~mag for MT91~372 and MT91~696, respectively, by
using a different set of UV photometry. If we had adopted these
extinctions, it would have increased the resulting distances to these
systems to 1.39 and 1.34~kpc, respectively. The overall weighted mean
distance would have risen slightly to $1.36\pm0.10$~kpc. This small
systematic difference is within the uncertainty of our
measurement, but it underscores the sensitivity of the distance
measurement to the adopted correction for interstellar extinction.

The level of agreement between our four distance measurements is
remarkable given the stated (perhaps overly generous) uncertainties
stemming primarily from uncertainties on effective temperatures and
interstellar reddenings. We regard this to be a coincidence given that
the analysis of each system has entailed independent measurements of
these key observable quantities arising from independent datasets.
The adopted photometry, reddenings, and stellar effective
temperatures are independently justified in each case, and we do not
believe that an artificial level of agreement has been imposed during
the analysis process.

\begin{figure*}
\centering
\includegraphics[width=\textwidth]{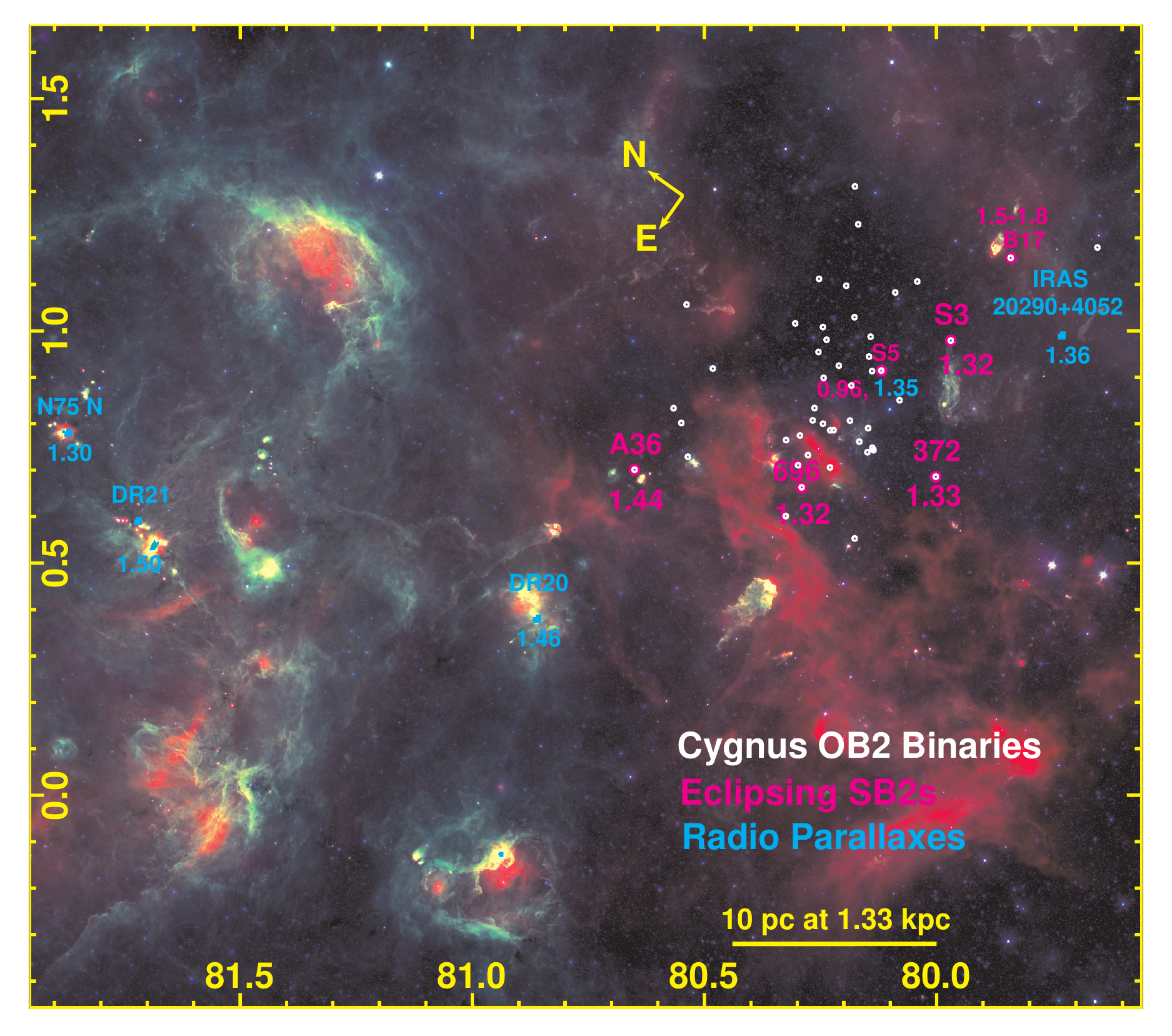}
\caption{\emph{Spitzer Space Telescope} 3.6 $\mu$m (cyan), 8.0 $\mu$m
  (green), and 24 $\mu$m (red) image of the Cygnus~X region.  White
  dots mark the locations of known massive binaries in Cygnus~OB2.
  Magenta labels mark eclipsing massive binaries and their derived
  distances in kpc.  Cyan labels denote radio parallax distance
  measurements toward masers from massive protostars
  \citep{Rygl12} and Schulte~5 \citep{Dzib12}. \label{color}}
\end{figure*}

Figure~\ref{color} displays a three-color representation of the
Cyg~OB2 vicinity, with blue, green, red depicting the
$Spitzer~Space~Telescope$ IRAC 4.5~$\mu$m, IRAC 8.0~$\mu$m, and MIPS
24 $\mu$m bands, respectively. Figure~\ref{color} shows the complex
nature of this region, including stellar photospheres (blue),
photo-dissociation regions at the edges of molecular clouds (diffuse
green tracing broad emission features arising from excited polycyclic
aromatic hydrocarbons, and hot dust (red).  White circles indicate the
48 known binary systems in Cyg~OB2 \citep{Kobulnicky14}, while magenta
labels highlight the known eclipsing binary systems and their
distances.  The yellow scale bar indicates the linear extent of 10 pc
at 1.33 kpc distance.  Blue labels mark four masers having radio
parallax distances \citep{Rygl12} and the massive eclipsing binary
Schulte 5 \citep{Dzib12}.

The maser distance measurements range from 1.3~--~1.5 kpc, in
agreement with most of the eclipsing binaries.  Given the large
angular extent of the Cygnus~X region (almost a degree!), the entire
star-forming complex containing Cyg~OB2 could reasonably stretch 100
pc or more along the line of sight, thereby encompassing the various
massive young stellar objects with maser parallaxes.  The eclipsing
binary CPR2002~B17 has a rather uncertain distance at 1.5~--~1.8 kpc
\citep{Stroud10} and could be part of an extended star-forming complex
somewhat further away than Cyg~OB2.  Schulte~5 is something of an
enigma. Both \citet{Linder2009} and \citet{Yasarsoy14} analyze the
joint light-curve and velocity-curve solution to compute distances of
0.93 and 0.97 kpc, respectively. While they both present evidence for
luminosity from a third body in this system, the magnitude of this
contribution is not known.  Nevertheless, a third or fourth body in
the Schulte~5 system \citep{Linder2009, Kennedy2010} would have to
contribute nearly as much light as the two eclipsing components in
order to raise the distance to the mean of other Cyg~OB2 eclipsing
systems.  The radio parallax measurement of Schulte~5 by
\citet{Dzib12} yield 1.3~--~1.4 kpc, in good agreement with the other
eclipsing binaries.  Given that there is no known population of
massive stars substantially in the foreground to Cyg~OB2, we think it
unlikely that Schulte~5 is a foreground object.  Neither is it likely
to be a runaway in the radial direction; its systemic velocity,
reported as -55 \kms\ \citep{Yasarsoy14}, would allow it to travel
$<$200 pc in the $\sim$3~--~4 Myr lifetime of the most massive stars.  A
refined understanding of the components of Schulte~5 are likely to
yield a larger distance in better agreement with the masers and
eclipsing systems.

\section{Conclusions}

We have presented a photometric and spectroscopic analysis of four
double-lined eclipsing binaries within the Cygnus~OB2 Association. The
joint analysis provides a measure of the temperatures, luminosities,
and thereby, the distances to these systems.  We find distances of
1.33$\pm$0.17, 1.32$\pm$0.07, 1.44$\pm$0.18, and 1.32$\pm$0.13 kpc
toward MT91~372, MT91~696, CPR2002~A36, and Schulte~3, respectively.
We adopt a weighted mean of 1.33$\pm$0.06 kpc as the best estimate for
the distance to the Cygnus~OB2 Association.  These are the most direct
distance estimates to this touchstone region of massive star
formation, and they agree well with radio VLBI parallax measurements
toward masers in the star-forming clouds on the periphery of Cyg~OB2.
Having a secure distance toward the stellar population of Cyg~OB2
allows more certainty in the luminosities, mass loss rates, and other
fundamental stellar quantities measured for the massive stars
therein. Within several years we anticipate parallaxes from the {\tt
  Gaia} space mission providing even more precise distances toward
Cygnus~X constituents.  Although {\tt Gaia's} precision is likely to
be limited in crowded regions near the Galactic Plane, refined
post-mission analysis of the data may yield a parallax distance to a
prominent region such as Cygnus OB2. It is our hope that this
eclipsing binary distance may serve as a benchmark for such an attempt.

\section*{Acknowledgements}

We would like to acknowledge the time, effort, and thoroughness of the
anonymous referee in assisting us to make this a stronger
manuscript. We thank Anirban Bhattacharjee for obtaining $V$-band
images of Schulte~3 at WIRO. We are very appreciative that Saida
Caballero-Nieves shared results from her adaptive optics imaging and
$HST$ Fine Guidance Sensor survey in advance of publication.  REU
summer students Eric Topel, Emily Rolen, and Katie Lester helped
obtain photometry of MT91~372.  Gregor Rauw encouraged us to pursue
the eclipsing binary distances. Andrej Pr{\v s}a provided advice on
the modeling of binary systems with {\tt PHEOBE}. Nick Wright provided
helpful comments pertaining to the introduction and massive star
population of Cyg~OB2. WIRO staff James Weger and Jerry Bucher enabled
this science through their efforts at the Observatory. Karen Kinemuchi
contributed observing assistance and data analysis during an early
phase of the Cyg~OB2 program.  We thank Robert D. Gehrz, John
A. Hackwell, and the State of Wyoming for their vision in 1975 to
build the Wyoming Infrared Observatory, without which this work would
not have been possible. This work was supported by the National
Science Foundation through Research Experience for Undergraduates
(REU) program grant AST 03-53760, through grant AST 03-07778, and
through grant AST 09-08239, and the support of the Wyoming NASA Space
Grant Consortium through grant \#NNX10A095H. D.~Kiminki thanks the
James Webb Space Telescope NIRCam project at the University of Arizona
for salary support. \textit{Facilities:} WIRO, RBO

%%%%%%%%%%%%%%%%%%%%%%%%%%%%%%%%%%%%%%%%%%%%%%%%%%




\newpage

\begin{appendix}

\setcounter{table}{0}
\renewcommand*\thetable{A.\arabic{table}}

\LongTables
\begin{deluxetable*}{lcrrrr}
\tablecolumns{6}
\tabletypesize{\scriptsize}
\tablewidth{0pc}
\tablecaption{Radial Velocities for MT91~372 \&\ MT91~696 \label{specdata.tab}}
\tablehead{
\colhead{} & 
\colhead{} & 
\colhead{$V_{r1}$} &
\colhead{$O_1-C_1$} &
\colhead{$V_{r2}$} &
\colhead{$O_2-C_2$} \\ 
\colhead{Date (HJD-2,400,000)} &
\colhead{$\phi$} &
\colhead{(\kms)} &
\colhead{(\kms)} &
\colhead{(\kms)} &
\colhead{(\kms)}}
\startdata
\multicolumn{6}{c}{MT91~372} \\
\hline
54,754.643.......................... & 0.633 &   79.8 (14.2) & -16.5 & -258.2 (14.4) & -19.0   \\
54,756.629.......................... & 0.124 &   19.4 (17.2) &   3.1 &  \nodata      & \nodata \\
54,758.616.......................... & 0.416 &  -72.0 (21.8) &   5.2 &  181.5 (36.2) & 28.5    \\
56,437.744.......................... & 0.237 & -139.2 (8.9)  &  -2.4 &  295.8 (11.5) & 5.9     \\
56,438.831.......................... & 0.725 &   96.8 (10.0) & -26.8 & -278.5 (17.6) & 23.8    \\
56,439.755.......................... & 0.140 & -120.1 (10.2) & -17.9 &  221.6 (13.8) & 10.0    \\
56,440.777.......................... & 0.199 &   71.1 (9.9)  &  -4.5 & -177.4 (12.3) & 14.8    \\
56,445.838.......................... & 0.871 &   77.9 (5.7)  &  -5.3 & -208.1 (9.5)  & 2.9     \\
56,446.752.......................... & 0.281 & -150.7 (9.0)  & -12.8 &  306.0 (12.4) & 13.8    \\
56,458.825.......................... & 0.701 &  127.0 (12.7) &   6.4 & -269.4 (15.6) & 26.0    \\
56,459.828.......................... & 0.151 & -122.3 (8.2)  & -14.1 &  242.6 (11.6) & 17.5    \\
56,466.865.......................... & 0.311 & -137.2 (8.1)  &  -4.6 &  273.7 (12.5) & -6.1    \\
\cutinhead{MT91~696}
54,699.741.......................... & 0.288 & -234.4 (9.2)  &   7.0 &  302.9 (14.8) &  -0.4 \\
54,700.772.......................... & 0.990 &   -4.9 (12.3) & -22.2 &   11.5 (15.5) &  30.6 \\
54,700.902.......................... & 0.078 & -139.4 (22.9) & -23.0 &  150.1 (27.6) &   2.7 \\
54,701.703.......................... & 0.623 &  187.9 (7.3)  &  12.4 & -233.9 (13.7) & -17.7 \\
54,701.910.......................... & 0.764 &  249.1 (9.4)  &  -0.4 & -319.9 (13.6) & -11.3 \\
54,724.617.......................... & 0.220 & -247.4 (5.1)  &  -3.5 &  300.0 (8.6)  &  -6.4 \\
54,724.728.......................... & 0.295 & -235.5 (8.8)  &   2.7 &  296.5 (10.9) &  -2.8 \\
54,724.824.......................... & 0.360 & -178.1 (19.4) &  12.6 &  225.5 (29.0) & -14.5 \\
54,725.609.......................... & 0.895 &  129.0 (6.1)  & -25.2 & -166.7 (8.3)  &  23.0 \\
54,725.736.......................... & 0.981 &    3.5 (8.8)  & -26.9 &   17.3 (10.6) &  52.8 \\
54,725.854.......................... & 0.061 & -114.5 (7.8)  & -21.7 &  122.5 (9.8)  &   4.4 \\
54,726.629.......................... & 0.189 &  137.4 (6.5)  &   4.1 & -181.4 (11.9) & -17.8 \\
54,726.743.......................... & 0.667 &  218.0 (7.3)  &   0.4 & -282.9 (9.8)  & -14.1 \\
54,726.842.......................... & 0.734 &  255.4 (16.8) &   6.3 & -318.7 (21.9) & -10.6 \\
54,729.629.......................... & 0.631 &  190.0 (6.6)  &   6.2 & -257.7 (10.1) & -31.1 \\
54,729.724.......................... & 0.696 &  242.2 (5.9)  &   6.2 & -302.2 (7.8)  & -10.1 \\
54,729.813.......................... & 0.757 &  255.0 (6.9)  &   4.8 & -297.6 (8.8)  &  11.7 \\
56,218.613.......................... & 0.111 & -162.4 (8.9)  &  -3.0 &  200.1 (11.5) &  -0.9 \\
56,222.568.......................... & 0.803 &  231.6 (5.9)  &  -5.1 & -280.3 (8.2)  &  12.2 \\
56,218.632.......................... & 0.124 & -167.3 (10.9) &   7.2 &  242.6 (13.7) &  22.8 \\
56,222.788.......................... & 0.953 &   62.6 (7.9)  & -11.1 &  -85.1 (11.8) &   4.3 \\
\cutinhead{CPR2002~A36}
54,403.610.......................... & 0.162 & -153.8 (7.1)  &  5.3     & 177.8 (4.2)  &  1.6   \\
54,403.702.......................... & 0.181 & -167.8 (6.7)  &  -2.3    & 193.8 (3.9)  &  2.8   \\
54,405.705.......................... & 0.610 & 9.9 (15.4)    &  -55.3   & -208.5 (9.1) &  -20.3 \\
54,406.632.......................... & 0.808 & 110.2 (4.8)   &  -0.1    & -257.5 (2.9) &  -9.0  \\
54,406.757.......................... & 0.835 & 104.7 (6.2)   &  2.2     & -243.7 (3.7) &  -10.7 \\
54,408.664.......................... & 0.243 & -185.7 (11.0) &  -12.0   & 234.9 (6.5)  &  20.7  \\
54,409.730.......................... & 0.471 & -39.9 (48.9)  &  16.0    & -14.5 (28.8) &  -50.6 \\
54,410.667.......................... & 0.671 & 90.8 (7.1)    &  -7.9    & -245.0 (4.2) &  -7.8  \\
54,641.827.......................... & 0.119 & -100.3 (10.2) &  38.4    & 121.2 (6.0)  &  -11.2 \\
54,642.754.......................... & 0.317 & -176.2 (7.1)  &  -15.7   & 211.5 (4.2)  &  16.8  \\
54,644.770.......................... & 0.748 & 116.9 (9.1)   &  1.0     & -252.1 (5.4) &  10.3  \\
54,645.834.......................... & 0.976 &  \nodata      &  \nodata & -25.6 (6.0)  &  47.4  \\
54,647.767.......................... & 0.389 & -148.2 (11.5) &  -25.1   & 115.4 (6.8)  &  -19.8 \\
54,669.843.......................... & 0.112 & -103.4 (8.7)  &  31.7    & 121.5 (5.1)  &  -2.0  \\
54,672.777.......................... & 0.739 & 109.5 (7.2)   &  -5.6    & -276.0 (4.2) &  -14.3 \\
54,673.875.......................... & 0.974 &  \nodata      &  \nodata & -62.3 (27.4) &  13.3  \\
54,696.684.......................... & 0.853 & 121.3 (6.8)   &  26.4    & -229.1 (4.0) &  -9.7  \\
54,696.964.......................... & 0.913 & 101.5 (8.0)   &  37.1    & -158.7 (4.7) &  -0.6  \\
54,697.659.......................... & 0.062 & -68.8 (19.5)  &  37.4    & 14.0 (11.5)  &  -41.6 \\
54,698.753.......................... & 0.296 & -196.2 (11.4) &  -29.5   & 206.3 (6.7)  &  1.0   \\
54,700.871.......................... & 0.749 & 103.6 (9.0)   &  -12.4   & -259.7 (5.3) &  2.7   \\
54,701.895.......................... & 0.968 & \nodata       &  \nodata & -78.4 (29.5) &  6.3   \\
\cutinhead{Schulte~3}
53,989.652.......................... & 0.643 &     42.1 (5.7)                     &  15.6  &  -224.4 (3.4)                   &   0.4 \\
53,989.774.......................... & 0.668 &     37.4 (4.1)                     &   3.4  &  -238.6 (3.4)                   &   5.8 \\
53,990.848.......................... & 0.895 &     16.3 (7.2)                     &  -0.8  &  -191.9 (4.0)                   &  -3.5 \\
54,286.904.......................... & 0.276 &   -145.4 (8.3)                     &  -7.2  &   201.2 (8.4)                   &  28.0 \\
54,341.802.......................... & 0.843 &    117.2 (12.3)\tablenotemark{a}   &  82.6  &  -217.4 (5.4)                   &  19.1 \\
54,342.724.......................... & 0.038 &      5.0 (104.8)\tablenotemark{a}  &  82.0  &    13.1 (46.0)                  &   8.1 \\
54,343.830.......................... & 0.271 &   -134.0 (6.0)\tablenotemark{a}    &   4.7  &   249.3 (2.6)\tablenotemark{a}  &  75.2 \\
54,344.761.......................... & 0.467 &    -21.1 (36.4)\tablenotemark{a}   &  46.2  &    11.0 (16.0)                  &   1.0 \\
54,345.781.......................... & 0.682 &     90.1 (9.8)\tablenotemark{a}    &  53.0  &  -251.2 (4.3)                   &   1.5 \\
54,346.787.......................... & 0.894 &     54.5 (25.4)\tablenotemark{a}   &  37.0  &  -170.1 (11.1)                  &  19.3 \\
54,347.755.......................... & 0.098 &    -19.0 (9.7)\tablenotemark{a}    &  89.8  &   101.4 (4.3)                   &  19.3 \\
54,348.840.......................... & 0.326 &   -116.8 (5.7)\tablenotemark{a}    &  13.1  &   220.4 (2.5)\tablenotemark{a}  &  66.8 \\
54,628.748.......................... & 0.305 &   -128.3 (4.7)                     &   6.1  &   172.4 (5.8)                   &   8.0 \\
54,628.880.......................... & 0.333 &   -134.2 (4.0)                     &  -5.8  &   162.7 (3.3)                   &  13.1 \\
54,629.866.......................... & 0.141 &    -49.7 (3.3)                     & -26.8  &  -126.7 (3.2)                   & -16.7 \\
54,630.739.......................... & 0.725 &     32.5 (4.4)                     & -11.1  &  -260.6 (2.6)                   &   8.7 \\
54,630.852.......................... & 0.749 &     40.2 (4.1)                     &  -4.8  &  -258.6 (2.8)                   &  13.5 \\
54,632.745.......................... & 0.147 &   -122.1 (4.5)                     &   4.7  &   155.4 (4.5)                   &  24.2 \\
54,633.741.......................... & 0.357 &   -140.0 (5.6)                     & -18.9  &   128.0 (5.9)                   &  -3.7 \\
54,633.859.......................... & 0.382 &   -120.0 (3.0)                     &  -8.3  &    82.9 (3.6)                   & -26.7 \\
54,642.717.......................... & 0.249 &   -128.9 (6.4)\tablenotemark{a}    &  10.9  &    258.2 (2.8)\tablenotemark{a} &  82.9 \\
54,642.878.......................... & 0.283 &   -143.6 (3.8)\tablenotemark{a}    &  -6.1  &    252.9 (1.7)\tablenotemark{a} &  81.0 \\
54,644.684.......................... & 0.663 &     91.2 (6.4)\tablenotemark{a}    &  58.6  &   -254.0 (2.8)                  & -13.3 \\
54,645.734.......................... & 0.884 &    101.5 (13.8)\tablenotemark{a}   &  80.3  &   -188.2 (6.1)                  &  11.6 \\
54,646.707.......................... & 0.089 &    -39.9 (8.9)\tablenotemark{a}    &  65.3  &    100.8 (3.9)                  &  28.7 \\ 
54,647.688.......................... & 0.296 &   -139.1 (8.0)\tablenotemark{a}    &  -3.2  &    229.0 (3.5)\tablenotemark{a} &  61.2 \\
54,648.902.......................... & 0.152 &    -13.9 (12.1)\tablenotemark{a}   &   2.6  &   -115.8 (5.3)                  &  10.0 \\
54,672.692.......................... & 0.165 &      0.3 (5.9)\tablenotemark{a}    &   9.6  &   -142.7 (2.6)                  &  -0.1 \\ 
56,218.402.......................... & 0.257 &   -124.8 (8.0)\tablenotemark{a}    &  14.6  &    223.8 (3.5)\tablenotemark{a} &  47.4 \\
56,219.341.......................... & 0.455 &    -56.1 (51.0)\tablenotemark{a}   &  17.5  &    -22.7 (22.4)                 & -48.4 \\
56,222.349.......................... & 0.089 &    -81.9 (7.9)\tablenotemark{a}    &  24.0  &    172.7 (3.5)\tablenotemark{a} &  98.1 
\enddata
\tablenotetext{a}{Observation not used in solution fit.}
\end{deluxetable*}

\clearpage

\twocolumngrid
\begin{deluxetable}{ccc}
\centering
\tabletypesize{\scriptsize}
\tablewidth{0pc}
\tablecaption{$V$-band photometry of MT91~372 \label{MT372phot.tab}}
\tablehead{
\colhead{HJD-2,400,000} &
\colhead{V} &
\colhead{$\sigma$\tablenotemark{a}} \\
\colhead{(days)  }   &
\colhead{(mag)  }   &
\colhead{(mag)  }  }
\startdata            
56,458.819.......................... & 14.983 & 0.011 \\
56,458.821.......................... & 14.991 & 0.012 \\
56,458.822.......................... & 14.936 & 0.011 \\
56,458.825.......................... & 14.955 & 0.011 \\
56,458.827.......................... & 14.962 & 0.011 \\
56,458.830.......................... & 14.977 & 0.011 \\
56,458.832.......................... & 14.993 & 0.011 \\
56,458.837.......................... & 14.986 & 0.011 \\
56,458.839.......................... & 14.994 & 0.011 \\
56,458.841.......................... & 14.964 & 0.011 
\enddata
\tablenotetext{a}{1 $\sigma$ error includes formal photon statistics
  uncertainties only; additional uncertainties at the level of
  0.01 mag are present as a result of uncorrected atmospheric
  transparency variations.}
\tablecomments{Table \ref{MT372phot.tab} is published in its entirety
  in the electronic edition of the Astrophysical Journal. A portion is
  shown here for guidance regarding its form and content.}
\end{deluxetable}

\begin{deluxetable}{ccc}
\centering
\tabletypesize{\scriptsize}
\tablewidth{0pc}
\tablecaption{$V$-band photometry of MT91~696 \label{MT696phot.tab}}
\tablehead{
\colhead{HJD-2,400,000} &
\colhead{V} &
\colhead{$\sigma$\tablenotemark{a}} \\
\colhead{(days)  }   &
\colhead{(mag)  }   &
\colhead{(mag)  }  }
\startdata            
56,219.567.......................... &    12.066  &    0.002  \\     
56,219.569.......................... &    12.068  &    0.002  \\     
56,219.570.......................... &    12.064  &    0.002  \\     
56,219.572.......................... &    12.067  &    0.002  \\     
56,219.573.......................... &    12.067  &    0.002  \\     
56,219.589.......................... &    12.065  &    0.002  \\     
56,219.590.......................... &    12.069  &    0.002  \\     
56,219.592.......................... &    12.063  &    0.002  \\     
56,219.593.......................... &    12.067  &    0.002  \\     
56,219.595.......................... &    12.069  &    0.002 
\enddata
\tablenotetext{a}{1 $\sigma$ error includes formal photon statistics
  uncertainties only; additional uncertainties at the level of
  0.01 mag are present as a result of uncorrected atmospheric
  transparency variations.}
\tablecomments{MT91~696 photometry includes the contribution from the blended 0\farcs02 tertiary
MT91~696c discussed in the text. Table \ref{MT696phot.tab} is published in its entirety
  in the electronic edition of the Astrophysical Journal. A portion is
  shown here for guidance regarding its form and content.}
\end{deluxetable}

\begin{deluxetable}{ccc}
\centering
\tabletypesize{\scriptsize}
\tablewidth{0pc}
\tablecaption{$V$-band photometry of CPR2002~A36 \label{A36phot.tab}}
\tablehead{
\colhead{HJD-2,400,000} &
\colhead{V} &
\colhead{$\sigma$\tablenotemark{a}} \\
\colhead{(days)  }   &
\colhead{(mag)  }   &
\colhead{(mag)  }  }
\startdata            
56,271.541.......................... & 11.421 & 0.002 \\
56,271.542.......................... & 11.432 & 0.002 \\
56,271.543.......................... & 11.428 & 0.002 \\
56,271.648.......................... & 11.447 & 0.002 \\
56,271.649.......................... & 11.434 & 0.002 \\
56,271.649.......................... & 11.429 & 0.002 \\
56,271.650.......................... & 11.435 & 0.002 \\
56,271.650.......................... & 11.428 & 0.002 \\
56,271.651.......................... & 11.430 & 0.002 \\
56,271.651.......................... & 11.439 & 0.002      
\enddata
\tablenotetext{a}{1 $\sigma$ error includes formal photon statistics
  uncertainties only; additional uncertainties at the level of
  0.01 mag are present as a result of uncorrected atmospheric
  transparency variations.}
\tablecomments{Table \ref{A36phot.tab} is published in its entirety
  in the electronic edition of the Astrophysical Journal. A portion is
  shown here for guidance regarding its form and content.}
\end{deluxetable} 

\begin{deluxetable}{ccc}
\centering
\tabletypesize{\scriptsize}
\tablewidth{0pc}
\tablecaption{$V$-band photometry of Schulte~3 \label{S3phot.tab}}
\tablehead{
\colhead{HJD-2,400,000} &
\colhead{V} &
\colhead{$\sigma$\tablenotemark{a}} \\
\colhead{(days)  }   &
\colhead{(mag)  }   &
\colhead{(mag)  }  }
\startdata            
56,218.586.......................... &      10.194   &  0.001   \\   
56,218.587.......................... &      10.182   &  0.001   \\   
56,218.587.......................... &      10.169   &  0.001   \\   
56,218.588.......................... &      10.168   &  0.001   \\   
56,218.588.......................... &      10.171   &  0.001   \\   
56,218.589.......................... &      10.175   &  0.001   \\   
56,218.589.......................... &      10.181   &  0.001   \\   
56,218.590.......................... &      10.156   &  0.001   \\   
56,218.590.......................... &      10.174   &  0.001   \\   
56,218.591.......................... &      10.149   &  0.001     
\enddata
\tablenotetext{a}{1 $\sigma$ error includes formal photon statistics
  uncertainties only; additional uncertainties at the level of
  0.01 mag are present as a result of uncorrected atmospheric
  transparency variations.}
\tablecomments{Schulte~3 photometry includes the contribution from the
  blended tertiary Schulte~3C discussed in the text.  Table
  \ref{S3phot.tab} is published in its entirety in the electronic
  edition of the Astrophysical Journal. A portion is shown here for
  guidance regarding its form and content.}
\end{deluxetable}

\end{appendix}
\end{document}